\documentclass{article}
\usepackage[utf8]{inputenc}
\usepackage[style = ieee]{biblatex}
\usepackage{amsmath}  
\usepackage{float}
\usepackage[margin=1in]{geometry}
\usepackage{caption}
\usepackage{hanging}

\usepackage{amssymb}
\usepackage{graphics}
\usepackage{graphicx}
\usepackage{dirtytalk}
\usepackage{csquotes}
\usepackage[english]{babel}
\usepackage{amsthm}
\usepackage{listings}
\usepackage{mathrsfs}
\usepackage{xcolor}

\title{Behavior of a Chiral Condensate Around Astrophysical-Mass Schwarzschild and Reissner-Nordstr\"om Black Holes}
\author{Ross DeMott, Alex Flournoy}
\date{\textit{Department of Physics, Colorado School of Mines, Golden, CO 80401, USA.} \\ \vspace{3 mm} April 2023}

\begin{document}

\maketitle

\begin{abstract}
\noindent In this work, we develop a perturbative method to describe the behavior of a chiral condensate around a spherical black hole whose mass is astrophysically realistic. We use the inverse mass as the expansion parameter for our perturbative series. We test this perturbative method in the case of a Schwarzschild black hole, and we find that it agrees well with previous numerical results. For an astrophysical-mass Schwarzschild black hole, the leading order contribution to the condensate is much larger (in most of space) than the next-to-leading order contribution, providing further evidence for the validity of the perturbative approach. The size of the bubble of restored chiral symmetry is directly proportional to the size of the black hole.
\newline
Next, we apply this perturbative method to a Reissner-Nordstr\"om (RN) black hole. We find that, as the charge-to-mass ratio increases, the bubble of restored chiral symmetry becomes larger relative to the black hole. This effect is particularly pronounced for near-extremal RN black holes. The case of an extremal RN black hole provides an interesting counterexample to the standard thermal explanation for the formation of a bubble of restored chiral symmetry around a black hole. 
\end{abstract}

\section{Introduction}
In physics, spontaneous symmetry breaking is a critical tool, used to explain how fundamentally symmetric theories can give rise to asymmetric low-energy effective theories. Most of the familiar examples of spontaneous symmetry breaking come from theories formulated in Minkowski spacetime, but it is also possible for spontaneous symmetry breaking to occur in theories on curved backgrounds \cite{hawking1981interacting, PhysRevD.32.1333, PhysRevLett.67.1975, COLEMAN1992175}.
\vspace{2 mm}
\newline The QCD phase diagram is a matter of great interest to contemporary particle physics. At high energies and/or high temperatures, QCD exhibits chiral symmetry. At low energies and/or temperatures, QCD spontaneously breaks chiral symmetry. If gravity is included in the theory, strong space-time curvature can also cause QCD matter to transition from a phase with spontaneously broken chiral symmetry to a phase with restored chiral symmetry \cite{FlachiTanaka2011-1, flachi2012chiral, Flachi2013-2, FlachiTanaka2011-chiralphasetransitions}. Such strong curvature can be found in the vicinity of black holes.
\vspace{2 mm}
\newline In Ref. \cite{FlachiTanaka2011-chiralphasetransitions}, the authors found that the chiral condensate approaches zero near the event horizon of a Schwarzschild black hole. Thus, the event horizon is surrounded by a ``bubble" of approximately restored chiral symmetry. As the mass of the black hole increases, the radius of this bubble (relative to the radius of the black hole) decreases. At the boundary of the bubble, a second-order phase transition takes place, in which the chiral-symmetric phase inside the bubble meets the chiral-broken phase in the wider universe.
\vspace{2 mm}
\newline 
Throughout this article, we restrict our discussion to astrophysical-mass black holes, whose masses are much larger than the Planck scale. Because of this, we can use the inverse mass (in Planck units) as a small expansion parameter. We use this perturbative approach to greatly simplify the equations of motion for the condensate $\sigma$. We apply this method to Schwarzschild black holes, and we find that the results thereby obtained are consistent with the numerical results obtained in Ref. \cite{FlachiTanaka2011-chiralphasetransitions}. Having confirmed the validity of the perturbative approach for Schwarzschild black holes, we apply the method to Reissner-Nordstr\"om black holes (which possess electric charge but no angular momentum). As far as we know, this is the first time that the behavior of a chiral condensate around a charged black hole has been investigated.
\vspace{2 mm}
\newline In flat spacetime, there is a critical temperature $T_{c}$ that marks the boundary between the phase with spontaneously broken chiral symmetry ($T < T_{c}$) and the phase with restored chiral symmetry ($T > T_{c}$) \cite{Fukushima_2011, FlachiTanaka2011-chiralphasetransitions}. At any point, the Tolman temperature gives the effective local temperature of spacetime \cite{FlachiTanaka2011-chiralphasetransitions}. As one approaches the event horizon of the black hole, the Tolman temperature increases without bound. Assuming that the asymptotic temperature is smaller than $T_{c}$, the Tolman temperature will cross $T_{c}$ at some finite radius outside the event horizon. Thus, one intuitively expects that the chiral condensate will transition from a spontaneously broken phase far away from the horizon to a restored phase close to the horizon. This is the usual argument explaining why chiral symmetry is restored close to the horizon \cite{FlachiTanaka2011-chiralphasetransitions}. However, we find that this intuitive argument fails for an extremal Reissner-Nordstr\"om black hole.
\vspace{2 mm}
\newline Throughout this article, we use Planck units, so that $c = G = \hbar = 4 \pi \hspace{0.5 mm} \epsilon_{0} = k_{B} = 1$.

\section{Effective Action for General Black Hole Space-Time}
In a Minkowski background, a general, spherically-symmetric black hole metric takes the form
\begin{equation} \label{gensphbhmetric}
ds^{2} = -f\left(r\right) \hspace{0.5 mm} d t^{2} + f\left(r\right)^{-1} \hspace{0.5 mm} d r^{2} + r^{2} \hspace{0.5 mm} d \Omega^{2} ,
\end{equation}
where $d \Omega^{2}$ is the line element on the two-sphere. Every black hole has a temperature $T_{\textrm{BH}}$. To account for thermal effects, we work in Euclidean time. The Euclidean time direction is periodic, and its period is equal to the inverse temperature $\beta$. In Euclidean time, we may write the metric as
\begin{equation}
ds^{2} = f\left(r\right) \hspace{0.5 mm} d t^{2} + f\left(r\right) ^{-1} \hspace{0.5 mm} d r^{2} + r^{2} \hspace{0.5 mm} d \Omega^{2} .
\end{equation}
We model baryonic matter by a four-fermion interaction with interaction strength $\lambda$. Let $N$ be the total number of different types of quarks, counting both flavors and colors. We may write the classical action as \cite{FlachiTanaka2011-chiralphasetransitions}
\begin{equation} \label{classicalaction}
S = \int \mathrm{d}^{4} x \hspace{0.5 mm} \left\{\overline{\psi} \hspace{0.5 mm} i \gamma^{\mu} \nabla_{\mu} \psi + \frac{\lambda}{2 N} \left(\overline{\psi} \psi\right)^{2} \right\} .
\end{equation}
(The astute reader may notice that there is no factor of $\sqrt{g}$ beside the volume element in Eqn. \ref{classicalaction}. This is because the determinant of the metric in Schwarzschild spacetime is identical to that in Minkowski spacetime.) We define the condensate $\sigma$ as
\begin{equation} \label{sigmadef}
\sigma = -\frac{\lambda}{N} \hspace{0.5 mm} \langle \overline{\psi} \psi \rangle .
\end{equation}
Let us introduce the new variables $\Sigma_{\pm}$, which are defined by
\begin{equation} \label{Sigmapmdef}
\Sigma_{\pm} = \sigma^{2} \pm \sqrt{f} \frac{d \sigma}{d r} .
\end{equation}
Note that $f$ is short for $f\left(r\right)$. Let us define $\omega_{\nu}\left(u\right)$ as
\begin{equation}
\omega_{\nu}\left(u\right) = \sum_{n = 1}^{\infty} \left(-1\right)^{n} n^{- \nu} K_{\nu}\left(n \beta u\right) .
\end{equation}
To approximate the effective action, we use a resummed heat kernel expansion. To second-order in the resummed heat kernel expansion, we may write the effective action $\Gamma\left[\sigma\right]$ as \cite{FlachiTanaka2011-chiralphasetransitions, FlachiTanaka2011-1, flachi2012chiral}
\begin{equation} \label{effactlagrange}
\Gamma\left[\sigma\right] = \beta \int \mathrm{d}^{3} \vec{x} \hspace{1.0 mm} \mathcal{L} ,
\end{equation}
where $\vec{x}$ are the spatial coordinates. The Lagrangian density $\mathcal{L}$ is given by \cite{FlachiTanaka2011-chiralphasetransitions, FlachiTanaka2011-1, flachi2012chiral} 
\begin{equation} \label{lagdensdecomp}
\mathcal{L} = - \frac{\sigma^{2}}{2 \lambda} + \frac{f^{-2}}{32 \pi ^{2}} \left(\mathcal{L}_{+} + \mathcal{L}_{-}\right) + \frac{1}{32 \pi^{2}} \delta \mathcal{L} .
\end{equation}
The Lagrangian pieces $\mathcal{L}_{\pm}$ and $\delta \mathcal{L}$ are given by
\begin{equation} \label{Lpmdef}
\mathcal{L}_{\pm} = \frac{3}{4} \Sigma_{\pm}^{2} - \left(\frac{1}{2} \Sigma_{\pm}^{2} + a_{\pm}\right) \ln\left(\frac{f \Sigma_{\pm}}{\ell^{2}}\right) + \frac{16 \Sigma_{\pm}}{f \beta^{2}} \hspace{0.5 mm} \omega_{2}\left(\sqrt{f \Sigma_{\pm}}\right) + 4 a_{\pm} \omega_{0}\left(\sqrt{f \Sigma_{\pm}}\right)
\end{equation}
and
\begin{equation} \label{deltaLdef}
\delta \mathcal{L} = \left(\Sigma_{+}^{2} + \Sigma_{-}^{2}\right) \frac{\ln f}{2} - \frac{\left(\Sigma_{+} + \Sigma_{-}\right)}{12 f} \left\{\left(\frac{d f}{d r}\right)^{2} - 2 f \frac{d^{2} f}{d r^{2}} + \frac{4 f}{r} \frac{d f}{d r}\right\} .
\end{equation}
With a little algebra, we may rewrite $\delta \mathcal{L}$ as
\begin{equation}
\delta \mathcal{L} = \left(\sigma^{4} + f \left(\frac{d \sigma}{d r}\right)^{2}\right) \ln f - \frac{\sigma^{2}}{6 f} \left\{\left(\frac{d f}{d r}\right)^{2} - 2 f \frac{d^{2} f}{d r^{2}} + \frac{4 f}{r} \frac{d f}{d r}\right\} .
\end{equation}
Let $\Delta_{\textrm{Eucl}}$ be the Laplacian for three-dimensional Euclidean space. We may write $a_{\pm}$ as (see Appendix A for derivation)
\begin{equation}
a_{\pm} = a_{\textrm{sp}} + \frac{f\left(r\right)^{2}}{6} \Delta_{\textrm{Eucl}}\left(f \Sigma_{\pm}\right) ,
\end{equation}
where the pure space-time contribution $a_{\textrm{sp}}$ is given by
\begin{align} \label{apmeqn}
\begin{split}
a_{\textrm{sp}} = \hspace{1.0 mm} & \frac{1}{180} \left\{ \frac{2 f\left(r\right)^{2}}{r^{4}} - \frac{8 f\left(r\right)^{3}}{r^{4}} + \frac{6 f\left(r\right)^{4}}{r^{4}} + \frac{8 f\left(r\right)^{2}}{r^{3}} f^{\prime}\left(r\right) - \frac{12 f\left(r\right)^{3}}{r^{3}} f^{\prime}\left(r\right) + \frac{6 f\left(r\right)^{2}}{r^{2}} f^{\prime}\left(r\right)^{2} - \frac{4 f\left(r\right)^{2}}{r^{2}} f^{\prime \prime}\left(r\right) \right. \\
& \left. + \frac{6 f\left(r\right)^{3}}{r^{2}} f^{\prime \prime}\left(r\right) - \frac{6 f\left(r\right)^{2}}{r} f^{\prime}\left(r\right) f^{\prime \prime}\left(r\right) + \frac{3}{2} f\left(r\right)^{2} f^{\prime \prime}\left(r\right)^{2} - \frac{6 f\left(r\right)^{3}}{r} f^{\left(3\right)}\left(r\right) - f\left(r\right)^{2} f^{\prime}\left(r\right) f^{\left(3\right)}\left(r\right) \right. \\
& \left. - 2 f\left(r\right)^{3} f^{\left(4\right)}\left(r\right) \right\} .
\end{split}
\end{align}
We assume the black hole is of astrophysical size, which means that $\beta$ is very large. Therefore, let us make the following assumptions:
\begin{equation} \label{assumplus}
\beta \sqrt{f \Sigma_{+}} \gg 1 ,
\end{equation}
\begin{equation} \label{assumminus}
\beta \sqrt{f \Sigma_{-}} \gg 1 .
\end{equation}
Later, we will verify that these assumptions are satisfied in most of space. The Bessel functions $K_{\nu}\left(x\right)$ decay exponentially for large values of $x$. Therefore, if Eqns. \ref{assumplus} and \ref{assumminus} are satisfied, the Lagrangian terms involving $\omega_{0}\left(\sqrt{f \Sigma_{\pm}}\right)$ and $\omega_{2}\left(\sqrt{f \Sigma_{\pm}}\right)$ are exponentially suppressed. Hence, we may simplify the Lagrangian pieces $\mathcal{L}_{\pm}$ to
\begin{equation}
\mathcal{L}_{\pm} = \frac{3}{4} \Sigma_{\pm}^{2} - \left(\frac{1}{2} \Sigma_{\pm}^{2} + a_{\pm}\right) \ln\left(\frac{f \Sigma_{\pm}}{\ell^{2}}\right) .
\end{equation}
With some algebraic manipulation, we may write the sum $\mathcal{L}_{+} + \mathcal{L}_{-}$ as
\begin{align} \label{Lpmsumschwarz}
\begin{split}
\mathcal{L}_{+} + \mathcal{L}_{-} = & \left(\frac{3}{2} - \ln\left(\frac{f}{\ell^{2}}\right)\right) \left(\sigma^{4} + f \left(\frac{d \sigma}{d r}\right)^{2}\right) - \frac{1}{2} \left(\Sigma_{+}^{2} \ln \Sigma_{+} + \Sigma_{-}^{2} \ln \Sigma_{-}\right) - \frac{f\left(r\right)^{2}}{3} \Delta_{\textrm{Eucl}}\left(f \sigma^{2}\right) \ln\left(\frac{f}{\ell^{2}}\right) \\
& - a_{\textrm{sp}} \ln\left(\sigma^{4} - f \left(\frac{d \sigma}{d r}\right)^{2}\right) - \frac{f\left(r\right)^{2}}{6} \Delta_{\textrm{Eucl}}\left(f \Sigma_{+}\right) \ln \Sigma_{+} - \frac{f\left(r\right)^{2}}{6} \Delta_{\textrm{Eucl}}\left(f \Sigma_{-}\right) \ln \Sigma_{-} .
\end{split}
\end{align}
(Note that we have omitted a term that does not depend on $\sigma$.) Recall that the effective action $\Gamma\left[\sigma\right]$ is given by
\begin{equation} \label{Gammaschwarz}
\Gamma\left[\sigma\right] = \beta \int \mathrm{d}^{3} \vec{x} \hspace{0.5 mm} \mathcal{L} 
\end{equation}
and that the total Lagrangian $\mathcal{L}$ is given by 
\begin{equation} \label{Lschwarz}
\mathcal{L} = - \frac{\sigma^{2}}{2 \lambda} + \frac{f^{-2}}{32 \pi ^{2}} \left(\mathcal{L}_{+} + \mathcal{L}_{-}\right) + \frac{1}{32 \pi^{2}} \delta \mathcal{L} .
\end{equation}
For convenience, let us introduce the quantities $\mathcal{A}$ and $\mathcal{B}$, which are defined as
\begin{equation}
\mathcal{A} = \left(\frac{3}{2} - \ln\left(\frac{f}{\ell^{2}}\right)\right) \left(\sigma^{4} + f \left(\frac{d \sigma}{d r}\right)^{2}\right) - \frac{1}{2} \left(\Sigma_{+}^{2} \ln \Sigma_{+} + \Sigma_{-}^{2} \ln \Sigma_{-}\right) - a_{\textrm{sp}} \ln\left(\sigma^{4} - f \left(\frac{d \sigma}{d r}\right)^{2}\right) ,
\end{equation}
\begin{equation}
\mathcal{B} = -\frac{1}{3} \Delta_{\textrm{Eucl}}\left(f \sigma^{2}\right) \ln\left(\frac{f}{\ell^{2}}\right) - \frac{1}{6} \Delta_{\textrm{Eucl}}\left(f \Sigma_{+}\right) \ln \Sigma_{+} - \frac{1}{6} \Delta_{\textrm{Eucl}}\left(f \Sigma_{-}\right) \ln \Sigma_{-} .
\end{equation}
From Eqns. \ref{Lpmsumschwarz} and \ref{Lschwarz}, we see that the Lagrangian may be written as
\begin{equation} \label{Lschwarznew}
\mathcal{L} = -\frac{\sigma^{2}}{2 \lambda} + \frac{f^{-2}}{32 \pi^{2}} \mathcal{A} + \frac{1}{32 \pi^{2}} \mathcal{B} + \frac{1}{32 \pi^{2}} \delta \mathcal{L} .
\end{equation}
Combining Eqns. \ref{Lpmsumschwarz}, \ref{Gammaschwarz}, and \ref{Lschwarznew}, we may write the effective action as
\begin{equation} \label{effactlaplacian}
\Gamma\left[\sigma\right] = \frac{\beta}{32 \pi^{2}} \int \mathrm{d}^{3} \vec{x} \left(-\frac{1}{3} \Delta_{\textrm{Eucl}}\left(f \sigma^{2}\right) \ln\left(\frac{f}{\ell^{2}}\right) - \frac{1}{6} \Delta_{\textrm{Eucl}}\left(f \Sigma_{+}\right) \ln \Sigma_{+} - \frac{1}{6} \Delta_{\textrm{Eucl}}\left(f \Sigma_{-}\right) \ln \Sigma_{-}\right) + \dots , 
\end{equation}
where $\dots$ represents terms that do not depend involve the Laplacian $\Delta_{\textrm{Eucl}}$. Using Green's first identity, we may rewrite Eqn. \ref{effactlaplacian} as
\begin{equation} \label{effactlap2}
\Gamma\left[\sigma\right] = \frac{\beta}{32 \pi^{2}} \int \mathrm{d}^{3} \vec{x} \left(\frac{1}{3} \nabla_{\textrm{Eucl}}\left(f \sigma^{2}\right) \cdot \frac{\nabla_{\textrm{Eucl}} f}{f} + \frac{1}{6} \nabla_{\textrm{Eucl}}\left(f \Sigma_{+}\right) \cdot \frac{\nabla \Sigma_{+}}{\Sigma_{+}} + \frac{1}{6} \nabla_{\textrm{Eucl}}\left(f \Sigma_{-}\right) \cdot \frac{\nabla \Sigma_{-}}{\Sigma_{-}}\right) + \dots .
\end{equation}
Taking advantage of spherical symmetry, we may rewrite Eqn. \ref{effactlap2} (after some algebra) as
\begin{equation}
\Gamma\left[\sigma\right] = \frac{\beta}{32 \pi^{2}} \int \mathrm{d}^{3} \vec{x} \left(\frac{2}{3} \frac{d f}{d r} \frac{d \left(\sigma^{2}\right)}{d r} + \frac{\sigma^{2}}{3 f} \left(\frac{d f}{d r}\right)^{2} + \frac{f}{6 \Sigma_{+}} \left(\frac{d \Sigma_{+}}{d r}\right)^{2} + \frac{f}{6 \Sigma_{-}} \left(\frac{d \Sigma_{-}}{d r}\right)^{2}\right) + \dots .
\end{equation}
Thus, we may rewrite the Lagrangian piece $\mathcal{B}$ as
\begin{equation}
\mathcal{B} = \frac{2}{3} \frac{d f}{d r} \frac{d \left(\sigma^{2}\right)}{d r} + \frac{\sigma^{2}}{3 f} \left(\frac{d f}{d r}\right)^{2} + \frac{f}{6 \Sigma_{+}} \left(\frac{d \Sigma_{+}}{d r}\right)^{2} + \frac{f}{6 \Sigma_{-}} \left(\frac{d \Sigma_{-}}{d r}\right)^{2} .
\end{equation}

\section{Large Schwarzschild Black Hole}

\subsection{Expansion in Powers of $M^{-1}$}
Let us consider a large Schwarzschild black hole with mass $M \gg 1$ in Planck units. The metric function $f\left(r\right)$ is given by
\begin{equation} \label{frschwarz}
f\left(r\right) = 1 - \frac{2 M}{r} .
\end{equation}
With this metric function, the quantity $a_{\textrm{sp}}$ takes the form \cite{Mathematica}
\begin{equation} \label{aspschextexp}
a_{\textrm{sp}} = \frac{4 M^{2}}{5 r^{6}} \left(1 - \frac{2 M}{r}\right)^{2} .
\end{equation}
So far, we have been working in Planck units. However, for an astrophysical black hole, the mass $M$ should be at least a few solar masses. One solar mass is around $10^{38}$ Planck masses. Therefore, the Schwarzschild radius of an astrophysical black hole will be at least $\sim 10^{38}$ Planck lengths. Thus, the metric tensor and its derivatives will hardly change at all if $r$ changes by a Planck length. Since the profile of $\sigma$ as a function of $r$ depends on the properties of space-time, it is reasonable to expect that $\sigma^{\prime}\left(r\right)$ will be very small. To obtain a better description, we rescale our unit of length by a factor of $M$. As a result, we have a new radial coordinate $R$, which is given by
\begin{equation} \label{Rdef}
R = \frac{r}{M} .
\end{equation}
Under this coordinate transformation, derivatives will transform as
\begin{equation}
\frac{d^{n}}{d r^{n}} \longrightarrow \frac{1}{M^{n}} \frac{d^{n}}{d R^{n}} .
\end{equation}
We may write the metric function $f\left(R\right)$ as
\begin{equation} \label{fRschwarz}
f\left(R\right) = 1 - \frac{2}{R} .
\end{equation}
Thus, in this new coordinate system, the event horizon is located at $R = 2$. Because Eqn. \ref{fRschwarz} does not contain any very large or very small coefficients, we see that the coordinate $R$ is well-suited to capturing the spatial variation in the properties of space-time. Therefore, it is also well-suited to capturing the spatial variation in $\sigma$. Plugging Eqn. \ref{Rdef} into Eqn. \ref{aspschextexp}, we find that
\begin{equation}
a_{\textrm{sp}} = \frac{4}{5 M^{4} R^{6}} \left(1 - \frac{2}{R}\right)^{2} .
\end{equation}
In Ref. \cite{FlachiTanaka2011-chiralphasetransitions}, the authors employed non-perturbative, numerical methods to solve for the condensate $\sigma\left(r\right)$. That work considered black holes whose mass was within a few orders of magnitude of the Planck mass. Since the black holes considered here are much larger, it is more convenient to expand the Lagrangian $\mathcal{L}$ in powers of $M^{-1}$. Keeping terms up to $M^{-2}$, we may write the Lagrangian pieces $\mathcal{A}$, $\mathcal{B}$, and $\delta \mathcal{L}$ as \cite{Mathematica}
\begin{equation} \label{AMexp}
\mathcal{A} = \left(\frac{3}{2} - \ln\left(\frac{f}{\ell^{2}}\right)\right) \left(\sigma^{4} + \frac{f}{M^{2}} \left(\frac{d \sigma}{d R}\right)^{2}\right) - \frac{1}{2} \left(\Sigma_{+}^{2} \ln \Sigma_{+} + \Sigma_{-}^{2} \ln \Sigma_{-}\right) ,
\end{equation}
\begin{equation} \label{BMexp}
\mathcal{B} = \frac{8 \sigma}{3 M^{2} R^{2}} \frac{d \sigma}{d R} + \frac{4 \sigma^{2}}{3 M^{2} R^{4} f} + \frac{f}{6 M^{2} \Sigma_{+}} \left(\frac{d \Sigma_{+}}{d R}\right)^{2} + \frac{f}{6 M^{2} \Sigma_{-}} \left(\frac{d \Sigma_{-}}{d R}\right)^{2} ,
\end{equation}
\begin{equation} \label{dLMexp}
\delta \mathcal{L} = \left(\sigma^{4} + \frac{f}{M^{2}} \left(\frac{d \sigma}{d R}\right)^{2}\right) \ln f - \frac{\sigma^{2}}{3 f} \left(\frac{8}{M^{2} R^{3}} - \frac{14}{M^{2} R^{4}}\right) .
\end{equation}
To further simplify Eqns. \ref{AMexp} and \ref{BMexp}, we must expand $\ln \Sigma_{\pm}$ and $\Sigma_{\pm}^{-1}$ in powers of $M^{-1}$. We may write $\Sigma_{\pm}$ as
\begin{equation}
\Sigma_{\pm} = \sigma^{2} \left(1 \pm \frac{\sqrt{f}}{M \sigma^{2}} \frac{d \sigma}{d R}\right) .
\end{equation}
Thus, we may write $\ln \Sigma_{\pm}$ and $\Sigma_{\pm}^{-1}$ as
\begin{equation}
\ln \Sigma_{\pm} = \ln\left(\sigma^{2}\right) + \ln\left(1 \pm \frac{\sqrt{f}}{M \sigma^{2}} \frac{d \sigma}{d R}\right) ,
\end{equation}
\begin{equation}
\Sigma_{\pm}^{-1} = \frac{1}{\sigma^{2}} \left(1 \pm \frac{\sqrt{f}}{M \sigma^{2}} \frac{d \sigma}{d R}\right)^{-1} .
\end{equation}
Using Taylor series (up to second order in $M^{-1}$), we may re-express $\ln \Sigma_{\pm}$ and $\Sigma_{\pm}^{-1}$ as
\begin{equation} \label{lnSigmataylor}
\ln \Sigma_{\pm} = \ln\left(\sigma^{2}\right) \pm \frac{\sqrt{f}}{M \sigma^{2}} \frac{d \sigma}{d R} - \frac{f}{2 M^{2} \sigma^{4}} \left(\frac{d \sigma}{d R}\right)^{2} ,
\end{equation}
\begin{equation} \label{Sigmainvtaylor}
\Sigma_{\pm}^{-1} = \frac{1}{\sigma^{2}} \left(1 \mp \frac{\sqrt{f}}{M \sigma^{2}} \frac{d \sigma}{d R} + \frac{f}{M^{2} \sigma^{4}} \left(\frac{d \sigma}{d R}\right)^{2}\right) .
\end{equation}
Next, we plug Eqns. \ref{lnSigmataylor} and \ref{Sigmainvtaylor} into Eqns. \ref{AMexp} and \ref{BMexp} (again keeping terms only up to $M^{-2}$) to obtain
\begin{equation}
\mathcal{A} = - \ln\left(\frac{f \sigma^{2}}{\ell^{2}}\right) \left(\sigma^{4} + \frac{f}{M^{2}} \left(\frac{d \sigma}{d R}\right)^{2}\right) + \frac{3}{2} \sigma^{4} ,
\end{equation}
\begin{equation}
\mathcal{B} = \frac{8 \sigma}{3 M^{2} R^{2}} \frac{d \sigma}{d R} + \frac{4 \sigma^{2}}{3 M^{2} R^{4} f} + \frac{4 f}{3 M^{2}} \left(\frac{d \sigma}{d R}\right)^{2} .
\end{equation}
Thus, we may write the total Lagrangian as
\begin{align} \label{Lagfinal}
\begin{split} 
\mathcal{L} = \hspace{1.0 mm} & -\frac{\sigma^{2}}{2 \lambda} + \frac{f^{-2}}{32 \pi^{2}} \left(- \ln\left(\frac{f \sigma^{2}}{\ell^{2}}\right) \left(\sigma^{4} + \frac{f}{M^{2}} \left(\frac{d \sigma}{d R}\right)^{2}\right) + \frac{3}{2} \sigma^{4}\right) \\
& + \frac{1}{32 \pi^{2}} \left(\frac{8 \sigma}{3 M^{2} R^{2}} \frac{d \sigma}{d R} - \frac{8 \sigma^{2}}{3 M^{2} R^{3} f} + \frac{6 \sigma^{2}}{M^{2} R^{4} f} + \frac{4 f}{3 M^{2}} \left(\frac{d \sigma}{d R}\right)^{2} + \left(\sigma^{4} + \frac{f}{M^{2}} \left(\frac{d \sigma}{d R}\right)^{2}\right) \ln f\right)
\end{split}
\end{align}
Since $\mathcal{L}$ is spherically symmetric, we may write the effective action as
\begin{equation}
\Gamma\left[\sigma\right] = \beta \int \mathrm{d} R \hspace{0.5 mm} R^{2} \mathcal{L} .
\end{equation}
Thus, we may write the Euler-Lagrange equation as
\begin{equation} \label{ELeqnssphere}
\frac{\partial \mathcal{L}}{\partial \sigma} - \frac{d}{d R} \left(\frac{\partial \mathcal{L}}{\partial \sigma^{\prime}}\right) - \frac{2}{R} \frac{\partial \mathcal{L}}{\partial \sigma^{\prime}} = 0 .
\end{equation}
Plugging Eqn. \ref{Lagfinal} into Eqn. \ref{ELeqnssphere}, we find that
\begin{align} \label{EqMot}
\begin{split}
& - \frac{\sigma}{\lambda} + \frac{f^{-2}}{32 \pi^{2}} \left(-4 \ln\left(\frac{f \sigma^{2}}{\ell^{2}}\right) \sigma^{3} + 4 \sigma^{3} - \frac{2 f}{M^{2} \sigma} \left(\frac{d \sigma}{d R}\right)^{2}\right) \\
& + \frac{1}{32 \pi^{2}} \left(\frac{8}{3 M^{2} R^{2}} \frac{d \sigma}{d R} - \frac{16 \sigma}{3 M^{2} R^{3} f} + \frac{12 \sigma}{M^{2} R^{4} f} + 4 \sigma^{3} \ln f\right) \\
& - \frac{1}{32 \pi^{2}} \frac{d}{d R}\left(-\ln\left(\frac{f \sigma^{2}}{\ell^{2}}\right) \frac{2 f^{-1}}{M^{2}} \frac{d \sigma}{d R} + \frac{8 \sigma}{3 M^{2} R^{2}} + \frac{8 f}{3 M^{2}} \frac{d \sigma}{d R} + \frac{2 f}{M^{2}} \frac{d \sigma}{d R} \ln f\right) \\
& - \frac{1}{16 \pi^{2} R} \left(-\ln\left(\frac{f \sigma^{2}}{\ell^{2}}\right) \frac{2 f^{-1}}{M^{2}} \frac{d \sigma}{d R} + \frac{8 \sigma}{3 M^{2} R^{2}} + \frac{8 f}{3 M^{2}} \frac{d \sigma}{d R} + \frac{2 f}{M^{2}} \frac{d \sigma}{d R} \ln f\right) = 0 .
\end{split}
\end{align}

\subsection{Leading Order Condensate Profile}
Let us expand the condensate $\sigma$ in powers of $M^{-2}$ as
\begin{equation} \label{sigmaexpansion}
\sigma\left(R\right) = \sigma_{0}\left(R\right) + \frac{\sigma_{2}\left(R\right)}{M^{2}} + \dots
\end{equation}
At leading order (meaning without any $M$ dependence), we may write the equation of motion (Eqn. \ref{EqMot}) as
\begin{equation} \label{EOM01}
-\frac{\sigma_{0}}{\lambda} + \frac{f^{-2}}{32 \pi^{2}} \left(-4 \ln\left(\frac{f \sigma_{0}^{2}}{\ell^{2}}\right) \sigma_{0}^{3} + 4 \sigma_{0}^{3}\right) + \frac{1}{32 \pi^{2}} \left(4 \sigma_{0}^{3} \ln f\right) = 0 .
\end{equation}
Clearly, there is trivial solution $\sigma_{0} = 0$ to Eqn. \ref{EOM01}. However, we expect $\sigma_{0}$ to be generally non-zero. Therefore, with some simple algebra, we may rewrite Eqn. \ref{EOM01} as
\begin{equation} \label{EOM02}
- \frac{1}{2 \lambda} + \frac{f^{-2}}{16 \pi^{2}} \left(\sigma_{0}^{2} - \sigma_{0}^{2} \ln\left(\frac{f \sigma_{0}^{2}}{\ell^{2}}\right) + f^{2} \ln\left(f\right) \sigma_{0}^{2}\right) = 0 .
\end{equation}
We may solve this equation to obtain (see Appendix B)
\begin{equation} \label{sigma0squaresol}
\sigma_{0}^{2} = -\frac{8 \pi^{2} f^{2}}{\lambda} \left[W_{-1}\left(-\frac{8 \pi^{2}}{e \lambda \ell^{2}} f^{3 - f^{2}}\right)\right]^{-1} .
\end{equation}
Eqn. \ref{sigma0squaresol} provides an expression for $\sigma_{0}^{2}$. Therefore, there are two possible expressions for $\sigma_{0}$, which differ from each other by a sign. Note that the Lagrangian $\mathcal{L}$ (Eqn. \ref{Lagfinal}) and the equation of motion (Eqn. \ref{EqMot}) are both symmetric under the transformation $\sigma \to - \sigma$. Therefore, it does not matter which expression for $\sigma_{0}$ we choose. For simplicity, we will choose the positive sign. Thus, we may write $\sigma_{0}$ as
\begin{equation} \label{sigma0sol}
\sigma_{0} = \sqrt{-\frac{8 \pi^{2} f^{2}}{\lambda} \left[W_{-1}\left(-\frac{8 \pi^{2}}{e \lambda \ell^{2}} f^{3 - f^{2}}\right)\right]^{-1}} .
\end{equation}
Below, we have graphed $\sigma_{0}\left(r\right)$ for several different combinations of $\lambda$ and $\ell$ \cite{Mathematica}.
\begin{figure}[H]
    \centering
    \includegraphics[scale=0.49]{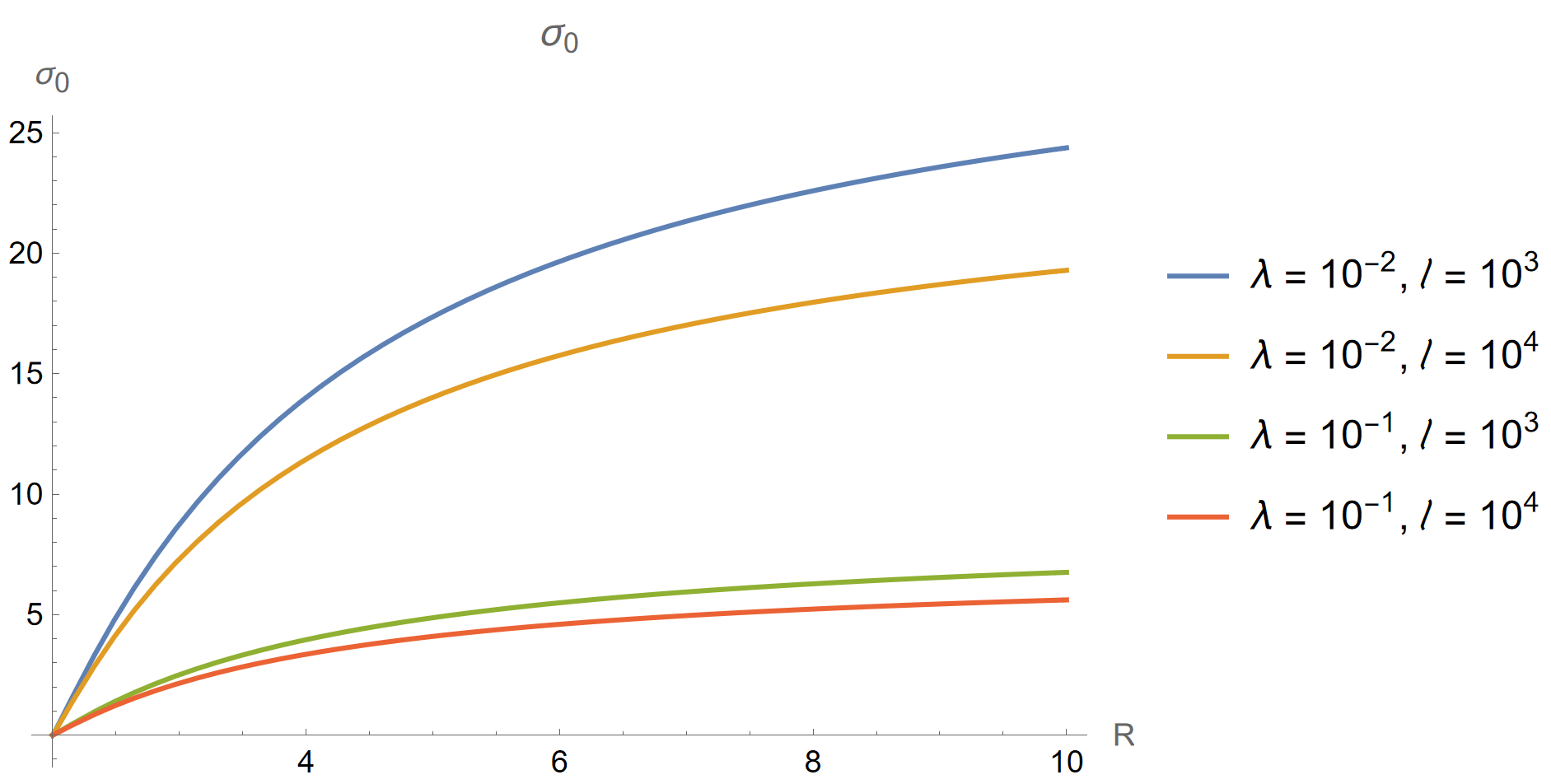}
    \caption{Graph of $\sigma_{0}\left(R\right)$ with respect to $R$. Note that the event horizon is at $R = 2$.}
    \label{fig:sigma0graph}
\end{figure}
\noindent As we will demonstrate in the next subsection, $\sigma_{0}$ provides a very close approximation to $\sigma$ almost everywhere outside the event horizon. Therefore, we would like to compare the results obtained in Figure \ref{fig:sigma0graph} to the results obtained in Ref. \cite{FlachiTanaka2011-chiralphasetransitions}. As in Ref. \cite{FlachiTanaka2011-chiralphasetransitions}, the condensate $\sigma$ asymptotically approaches a constant, non-zero value far from the event horizon, indicating a phase of spontaneously broken chiral symmetry. Close to the event horizon, $\sigma$ approaches zero, illustrating a phase of restored chiral symmetry.
\vspace{2 mm}
\newline In Ref. \cite{FlachiTanaka2011-chiralphasetransitions}, the authors work with black holes far smaller than the astrophysical-mass black holes considered here. They found that, as the mass of the black hole increases, the size of the bubble of restored chiral symmetry decreases (relative to the Schwarzschild radius of the black hole). For their black holes, the bubbles expanded out to $\sim 10 - 30$ Schwarzschild radii. In Fig. \ref{fig:sigma0graph}, the bubble extends out to $\sim 2 - 3$ Schwarzschild radii. Hence, we see that our results are consistent with the results of Ref. \cite{FlachiTanaka2011-chiralphasetransitions}, in that larger black holes create smaller bubbles of restored chiral symmetry (relative to their Schwarzschild radii). However, for very large black holes (like the astrophysical-mass black holes considered here), the size of the bubble scales almost linearly with the mass of the black hole. 

\subsection{Next-to-Leading Order Condensate Profile}
From Eqn. \ref{Lagfinal}, we see that $\mathcal{L}$ contains terms proportional to $\ln \sigma$. We wish to expand this term in powers of $M^{-2}$. To do this, we must first rewrite Eqn. \ref{sigmaexpansion} as
\begin{equation}
\sigma\left(r\right) = \sigma_{0}\left(R\right) \left(1 + \frac{\sigma_{2}\left(R\right)}{M^{2} \hspace{0.5 mm} \sigma_{0}\left(R\right)} + \dots\right) .
\end{equation}
Using the Mercator series for the natural logarithm, we may write $\ln \sigma$ as
\begin{equation}
\ln \sigma = \ln \sigma_{0} + \frac{\sigma_{2}}{M^{2} \hspace{0.5 mm} \sigma_{0}} - \frac{1}{2} \left(\frac{\sigma_{2}}{M^{2} \hspace{0.5 mm} \sigma_{0}}\right)^{2} + \dots
\end{equation}
At next-to-leading order (terms proportional to $M^{-2}$), we may write the equation of motion (Eqn. \ref{EqMot}) as
\begin{align} \label{sigma2eqmot}
\begin{split}
& - \frac{\sigma_{2}}{\lambda} + \frac{f^{-2}}{32 \pi^{2}} \left(-12 \ln\left(\frac{f \sigma_{0}^{2}}{\ell^{2}}\right) \sigma_{0}^{2} \sigma_{2} + 4 \sigma_{0}^{2} \sigma_{2} - \frac{2 f}{\sigma_{0}} \left(\frac{d \sigma_{0}}{d R}\right)^{2}\right) \\
& + \frac{1}{32 \pi^{2}} \left(\frac{8}{3 R^{2}} \frac{d \sigma_{0}}{d R} - \frac{16 \sigma_{0}}{3 R^{3} f} + \frac{12 \sigma_{0}}{R^{4} f} + 12 \sigma_{0}^{2} \sigma_{2} \ln f\right) \\
& - \frac{1}{32 \pi^{2}} \frac{d}{d R}\left(-2 f^{-1} \ln\left(\frac{f \sigma_{0}^{2}}{\ell^{2}}\right) \frac{d \sigma_{0}}{d R} + \frac{8 \sigma_{0}}{3 R^{2}} + \frac{8 f}{3} \frac{d \sigma_{0}}{d R} + 2 f \ln\left(f\right) \frac{d \sigma_{0}}{d R}\right) \\
& - \frac{1}{16 \pi^{2} R} \left(-2 f^{-1} \ln\left(\frac{f \sigma_{0}^{2}}{\ell^{2}}\right) \frac{d \sigma_{0}}{d R} + \frac{8 \sigma_{0}}{3 R^{2}} + \frac{8 f}{3} \frac{d \sigma_{0}}{d R} + 2 f \ln\left(f\right) \frac{d \sigma_{0}}{d R}\right) = 0 .
\end{split}
\end{align}
Solving Eqn. \ref{sigma2eqmot} for $\sigma_{2}$, we obtain
\begin{align} \label{sigma2sol}
\begin{split}
\sigma_{2} = & \frac{1}{32 \pi^{2}} \left\{-\frac{8}{3 R^{2}} \frac{d \sigma_{0}}{d R} + \frac{16 \sigma_{0}}{3 R^{3} f} - \frac{12 \sigma_{0}}{R^{4} f} + \frac{2 f^{-1}}{\sigma_{0}} \left(\frac{d \sigma_{0}}{d R}\right)^{2} \right. \\
& \left. + \frac{d}{d R}\left(-2 f^{-1} \ln\left(\frac{f \sigma_{0}^{2}}{\ell^{2}}\right) \frac{d \sigma_{0}}{d R} + \frac{8 \sigma_{0}}{3 R^{2}} + \frac{8 f}{3} \frac{d \sigma_{0}}{d R} + 2 f \ln\left(f\right) \frac{d \sigma_{0}}{d R}\right) \right. \\
& \left. + \frac{2}{R} \left(-2 f^{-1} \ln\left(\frac{f \sigma_{0}^{2}}{\ell^{2}}\right) \frac{d \sigma_{0}}{d R} + \frac{8 \sigma_{0}}{3 R^{2}} + \frac{8 f}{3} \frac{d \sigma_{0}}{d R} + 2 f \ln\left(f\right) \frac{d \sigma_{0}}{d R}\right)\right\} \\
& \times \left\{-\frac{1}{\lambda} + \frac{f^{-2}}{32 \pi^{2}} \left(-12 \ln\left(\frac{f \sigma_{0}^{2}}{\ell^{2}}\right) \sigma_{0}^{2} + 4 \sigma_{0}^{2} + 12 \sigma_{0}^{2} \hspace{0.5 mm} f^{2} \ln f\right)\right\}^{-1}
\end{split}
\end{align}
Because the explicit expression for $\sigma_{2}\left(R\right)$ is extremely complicated, we will not write it down here. Below, we have graphed $\sigma_{2}\left(R\right)$ for several different combinations of $\lambda$ and $\ell$ \cite{Mathematica}.
\begin{figure}[H]
    \centering
    \includegraphics[scale=0.445]{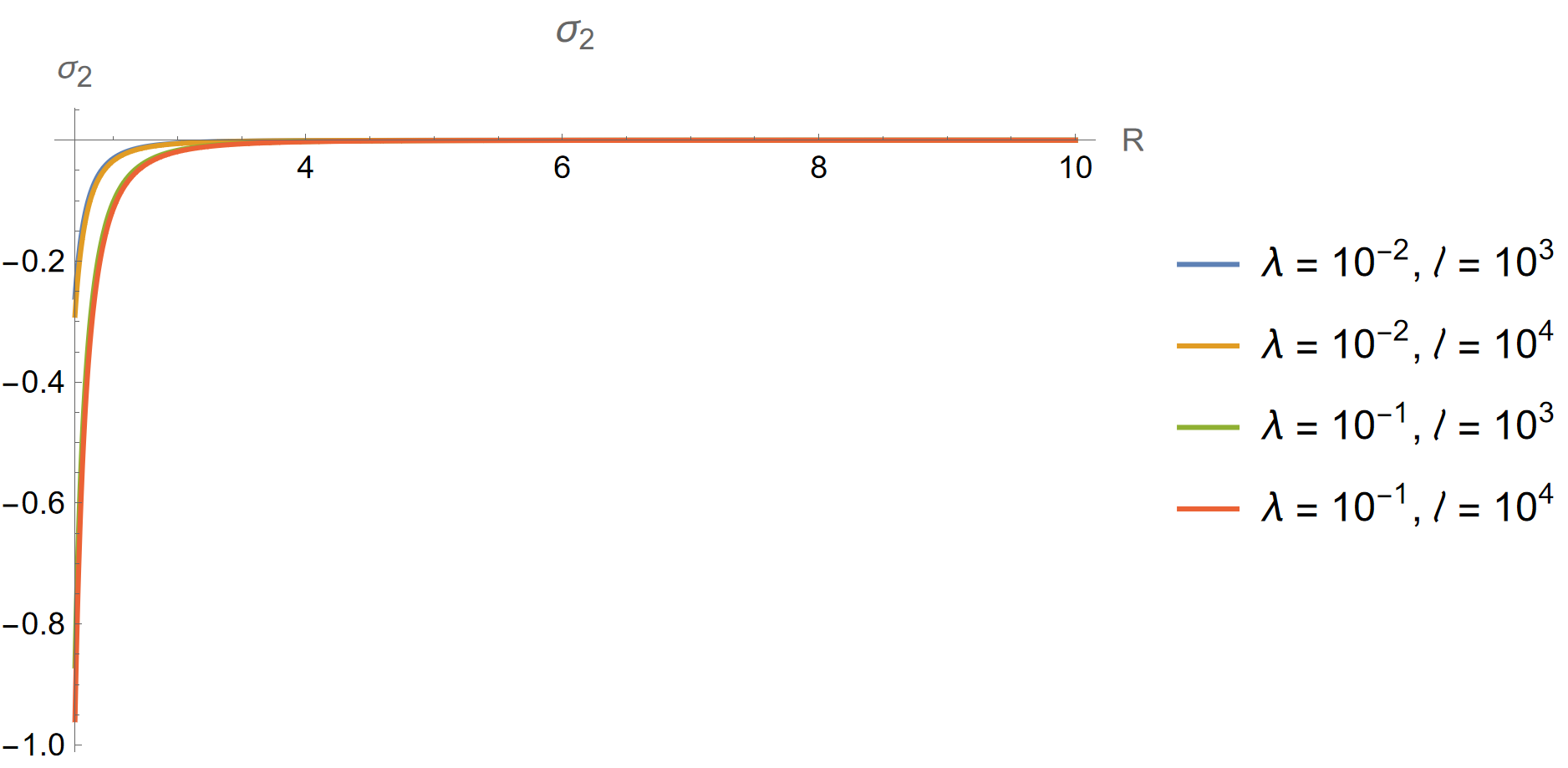}
    \caption{Graph of $\sigma_{2}\left(R\right)$ with respect to $R$.}
    \label{fig:sigma2graph}
\end{figure}
In the perturbative expansion for $\sigma$, $\sigma_{2}$ has a coefficient of $M^{-2}$. Since $\sigma_{2}$ is smaller than $\sigma_{0}$ almost everywhere outside the event horizon, the profile of $\sigma$ outside the horizon is dominated by the contribution from $\sigma_{0}$.

\subsection{Verification of the Perturbative Approach}
To obtain the above results for $\sigma_{0}$ and $\sigma_{2}$, we had to make three assumptions. First, we assumed that $\beta \sqrt{f \Sigma_{\pm}} \gg 1$ for both $\Sigma_{+}$ and $\Sigma_{-}$. Second, we assumed that $\sigma$ varies on scales far larger than the Planck scale. Lastly, we assumed that $\sigma$ can be expanded perturbatively in powers of $M^{-2}$. We wish to verify that these assumptions hold almost everywhere outside the event horizon.
\vspace{2 mm}
\newline We start by verifying the assumption $\beta \sqrt{f \Sigma_{\pm}} \gg 1$. As discussed earlier, an astrophysical black hole should have a mass of at least $\sim 10^{38}$ Planck masses. The inverse temperature $\beta$ is given by \cite{hawking1975particle, HawkingRad2}
\begin{equation}
\beta = 8 \pi M .
\end{equation}
Therefore, $\beta$ should be at least $\sim 10^{39}$. Hence, we see that the assumption is satisfied if $f \Sigma_{\pm} \gg 10^{-78}$. We may write $\Sigma_{\pm}$ as
\begin{equation}
\Sigma_{\pm} = \sigma^{2} \pm \frac{\sqrt{f}}{M} \frac{d \sigma}{d R} .
\end{equation}
Next, we use Eqn. \ref{sigmaexpansion} to expand $\Sigma_{\pm}$ in powers of $M^{-1}$. Keeping terms up to $M^{-2}$, we find that
\begin{equation}
\Sigma_{\pm} = \sigma_{0}^{2} + \frac{2}{M^{2}} \sigma_{0} \sigma_{2} \pm \frac{\sqrt{f}}{M} \frac{d \sigma_{0}}{d R} .
\end{equation}
Fortunately, we already have expressions for $\sigma_{0}$ and $\sigma_{2}$. Below, we graph the functions $f\left(R\right) \Sigma_{+}\left(R\right)$ and $f\left(R\right) \Sigma_{-}\left(R\right)$ for $M = 10^{39}$ Planck masses (a stellar-mass black hole) and $M = 10^{48}$ Planck masses (a very large supermassive black hole) \cite{Mathematica}.
\begin{figure}[H]
    \centering
    \includegraphics[scale=0.55]{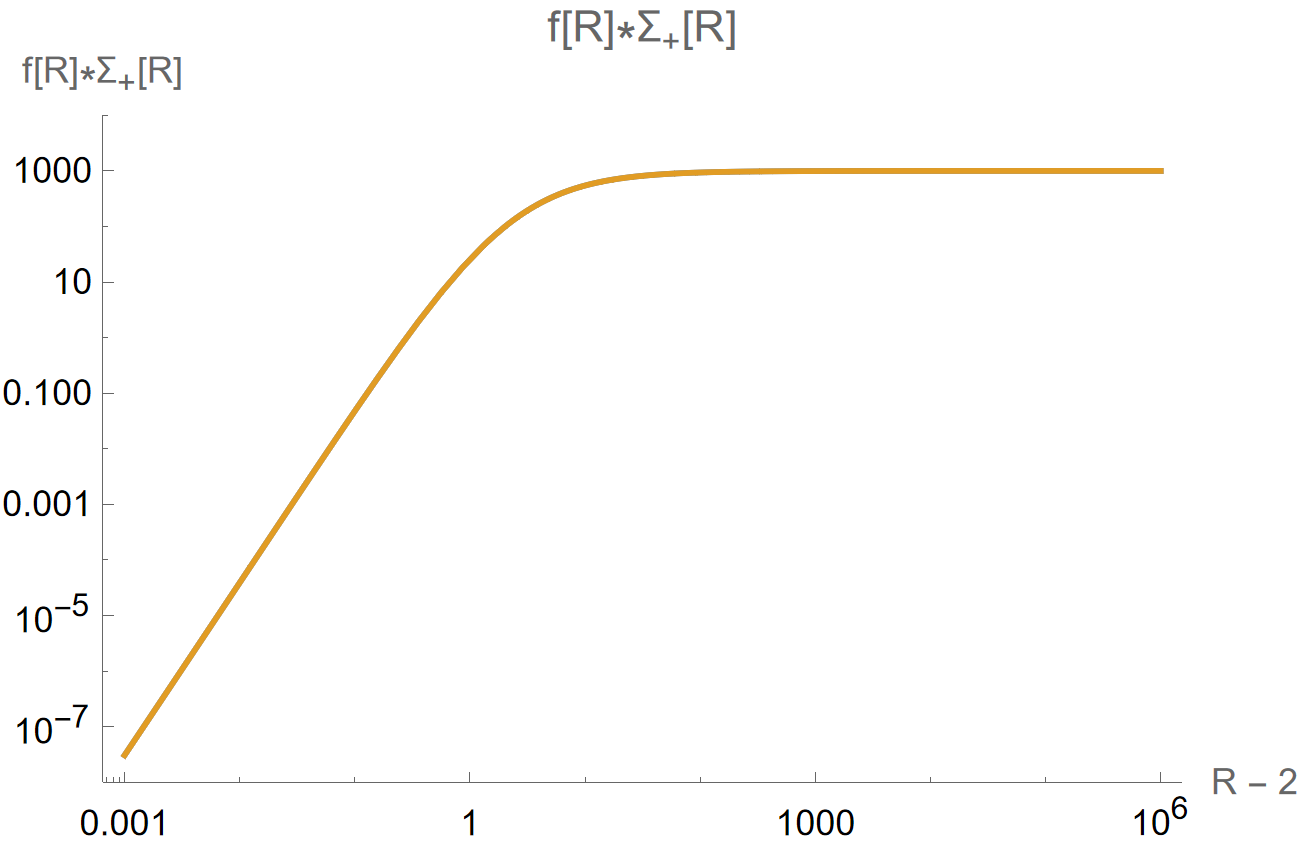}
    \caption{Graph of $f\left(R\right) \Sigma_{+}\left(R\right)$ with respect to $\left(R - 2\right)$, with $\lambda = 10^{-2}$ and $\ell = 10^{3}$. The curves for $M = 10^{39}$ Planck masses and $M = 10^{48}$ Planck masses are indistinguishable on the graph.}
    \label{fig:SigmaPlusGraph}
\end{figure}
\begin{figure}[H]
    \centering
    \includegraphics[scale=0.5]{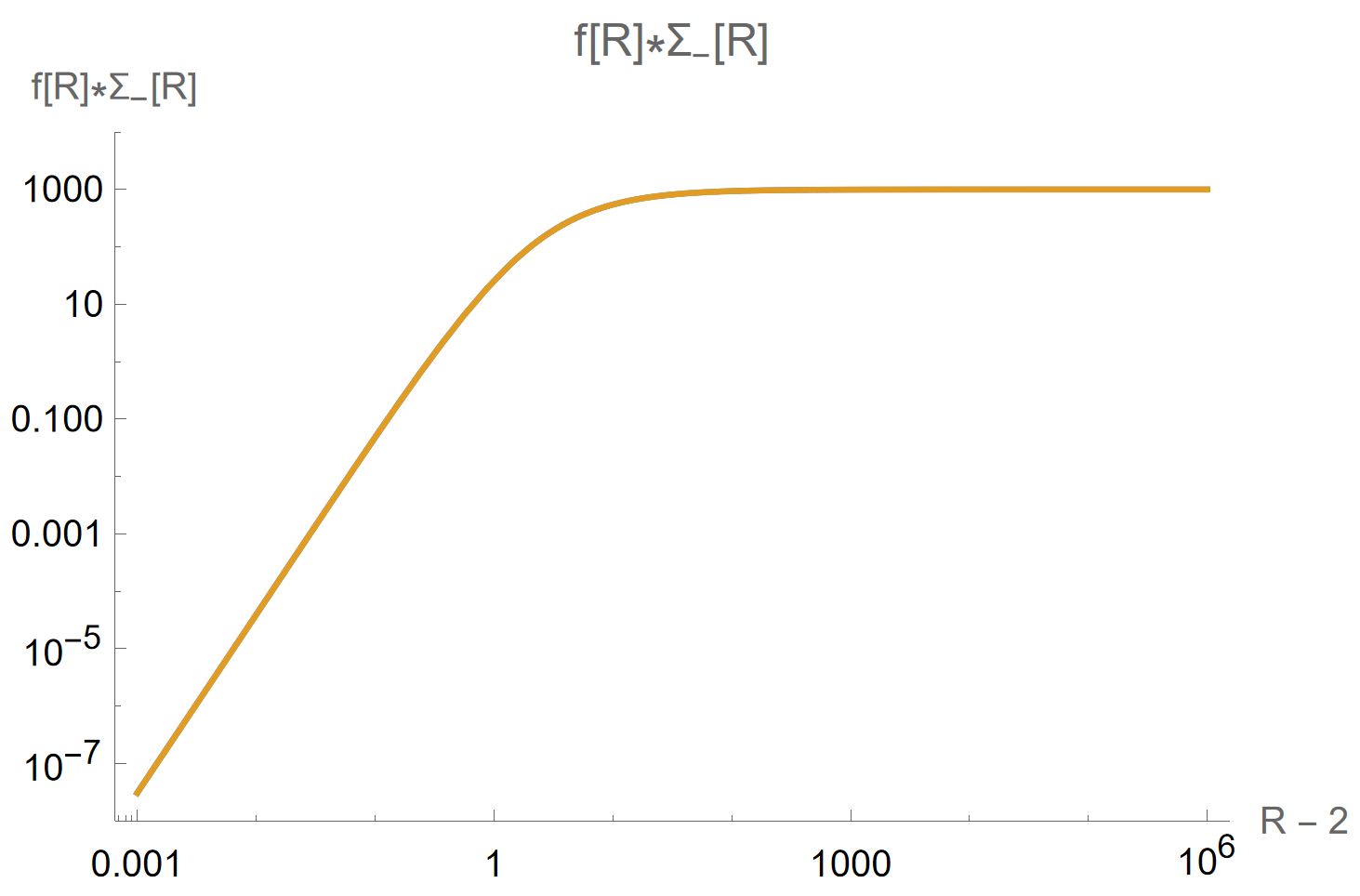}
    \caption{Graph of $f\left(R\right) \Sigma_{-}\left(R\right)$ with respect to $\left(R - 2\right)$, with $\lambda = 10^{-2}$ and $\ell = 10^{3}$. The curves for $M = 10^{39}$ Planck masses and $M = 10^{48}$ Planck masses are indistinguishable on the graph.}
    \label{fig:SigmaMinusGraph}
\end{figure}
\noindent From Figures \ref{fig:SigmaPlusGraph} and \ref{fig:SigmaMinusGraph}, we see that the assumption $\beta \sqrt{f \Sigma_{\pm}} \gg 1$ may be violated very close to the event horizon. However, the region where this occurs only extends out to $\left(R - 2\right) < 10^{-3}$, or less than one thousandth of a Schwarzschild radius outside the event horizon. Thus, the assumption $\beta \sqrt{f \Sigma_{\pm}} \gg 1$ holds almost everywhere outside the event horizon.
\vspace{2 mm}
\newline Next, we verify the assumption that $\sigma$ varies on scales far larger than the Planck scale. In terms of the original Planck-scale coordinate $r$, we may write this condition as
\begin{equation} \label{seccondorigr}
\frac{d \sigma}{d r} \ll 1 .
\end{equation}
Rewriting Eqn. \ref{seccondorigr} in terms of the rescaled coordinate $R$, we obtain
\begin{equation} \label{dsigmadRllM}
\frac{d \sigma}{d R} \ll M .
\end{equation}
Up to second-order in $M^{-1}$, we may write $\sigma$ as
\begin{equation}
\sigma\left(R\right) = \sigma_{0}\left(R\right) + \frac{\sigma_{2}\left(R\right)}{M^{2}} .
\end{equation}
Fortunately, we already have expressions for $\sigma_{0}$ and $\sigma_{2}$. Below, we graph the derivative of $\sigma\left(R\right)$ for $M = 10^{39}$ Planck masses and $M = 10^{48}$ Planck masses \cite{Mathematica}. We see that Eqn. \ref{dsigmadRllM} is satisfied everywhere outside the event horizon.
\begin{figure}[H]
    \centering
    \includegraphics[scale=0.65]{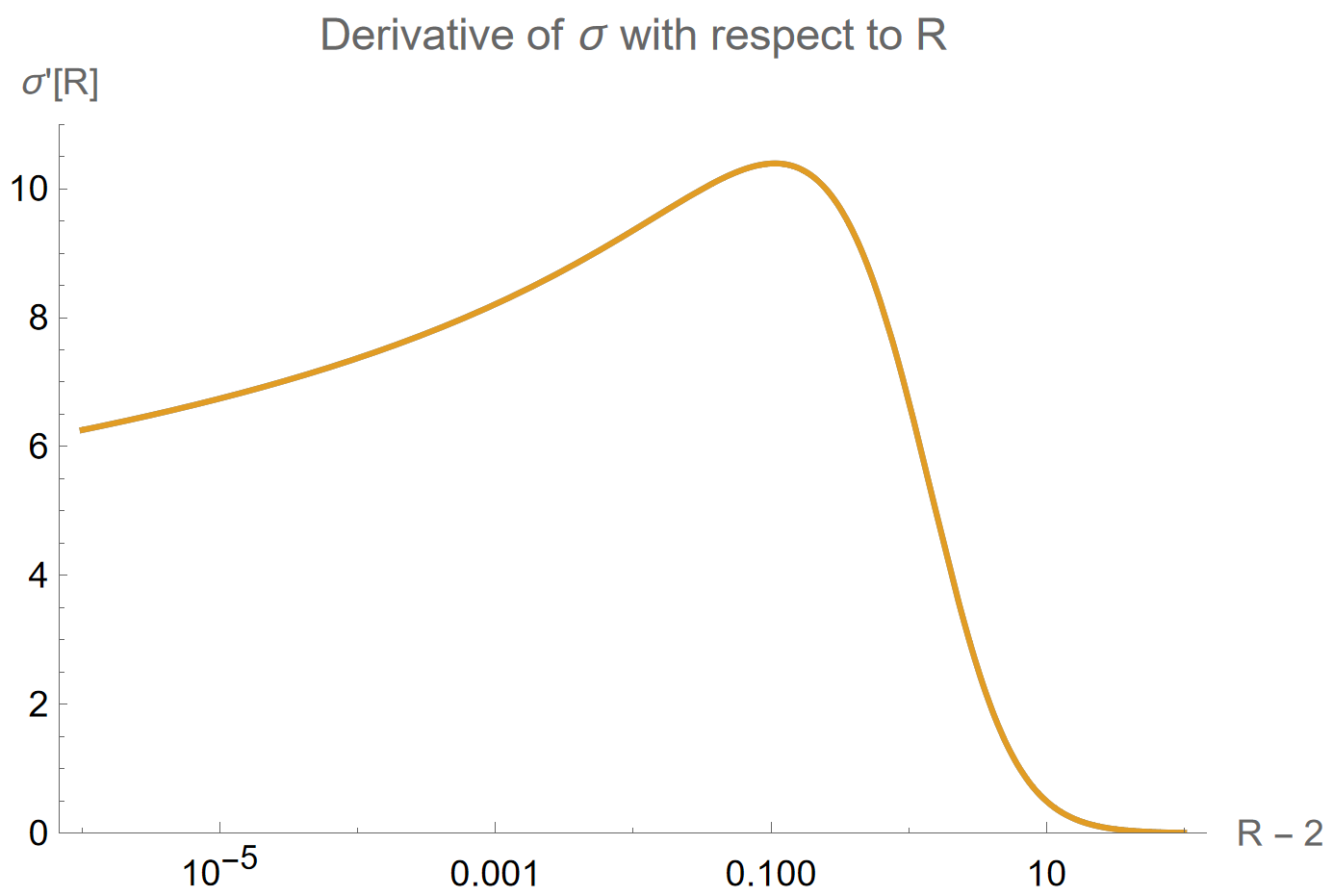}
    \caption{Graph of $\sigma^{\prime}\left(R\right)$ with respect to $\left(R - 2\right)$, with $\lambda = 10^{-2}$ and $\ell = 10^{3}$. The curves for $M = 10^{39}$ Planck masses and $M = 10^{48}$ Planck masses are indistinguishable on the graph.}
    \label{fig:SigmaDerivGraph}
\end{figure}
\noindent Finally, we verify the assumption that $\sigma$ can be expanded perturbatively in powers of $M^{-2}$, as in Eqn. \ref{sigmaexpansion}. This assumption holds if and only if $\lvert \sigma_{2}/M^{2} \rvert \ll \sigma_{0}$, or equivalently, $\lvert \sigma_{2} \rvert/\sigma_{0} \ll M^{2}$. From Figures \ref{fig:sigma0graph} and \ref{fig:sigma2graph}, it is easy to see that $\lvert \sigma_{2} \rvert < \sigma_{0}$ in most of the space outside the event horizon. Thus, $\lvert \sigma_{2} \rvert/\sigma_{0} \ll M^{2}$ almost everywhere outside the event horizon. Having verified all three assumptions, we conclude that the perturbative analysis is valid almost everywhere outside the event horizon.

\section{Large Reissner-Nordstr\"om Black Hole}
Let us consider a large Reissner-Nordstr\"om (RN) black hole with mass $M \gg 1$ in Planck units. We denote the charge of this black hole by $Q$. The event horizon radius $r_{+}$ is given by \cite{Misner:1973prb, HawkingRad2}
\begin{equation} \label{RNEHrad}
r_{+} = M + \sqrt{M^{2} - Q^{2}} .
\end{equation} 
The metric function $f\left(r\right)$ is given by \cite{Misner:1973prb}
\begin{equation}
f\left(r\right) = 1 - \frac{2 M}{r} + \frac{Q^{2}}{r^{2}} .
\end{equation}
We assume that the black hole is sub-extremal or extremal, so that $\lvert Q \rvert \leq M$ \cite{Misner:1973prb}. As in the Schwarzschild case, we assume that the event horizon radius sets the natural scale of variation for the condensate $\sigma$. Thus, we define a new radial coordinate $R$, which is given by
\begin{equation}
R = \frac{r}{M} .
\end{equation}
In this new coordinate system, we may write the metric function $f\left(R\right)$ as
\begin{equation}
f\left(R\right) = 1 - \frac{2}{R} + \frac{Q^{2}}{M^{2} R^{2}} .
\end{equation}
Let us expand the condensate $\sigma$ in powers of $M^{-2}$ as
\begin{equation}
\sigma\left(R\right) = \sigma_{0}\left(R\right) + \frac{\sigma_{2}\left(R\right)}{M^{2}} + \dots
\end{equation}
We may write $\sigma_{0}$ and $\sigma_{2}$ as (see Appendix C)
\begin{equation} \label{sigma0solRN}
\sigma_{0} = \sqrt{-\frac{8 \pi^{2} f^{2}}{\lambda} \left[W_{-1}\left(-\frac{8 \pi^{2}}{e \lambda \ell^{2}} f^{3 - f^{2}}\right)\right]^{-1}} ,
\end{equation}
\begin{align} \label{sigma2solRN}
\begin{split}
\sigma_{2} = & \frac{1}{32 \pi^{2}}\left\{\frac{2 f^{-1}}{\sigma_{0}} \left(\frac{d \sigma_{0}}{d R}\right)^{2} - \frac{8}{3 R^{2}} \left(1 - \frac{Q^{2}}{M^{2} R}\right) \frac{d \sigma_{0}}{d R} \right. \\
& \left. -\frac{8 \sigma_{0}}{3 R^{4} f} \left(1 - \frac{Q^{2}}{M^{2} R}\right)^{2} + \frac{2 \sigma_{0}}{3 f} \left(\frac{8}{R^{3}} - \frac{14}{R^{4}} + \frac{24 Q^{2}}{M^{2} R^{5}} - \frac{10 Q^{2}}{M^{2} R^{4}} - \frac{8 Q^{4}}{M^{4} R^{6}}\right) \right. \\
& \left. + \frac{d}{d R} \left[- 2 f^{-1} \hspace{0.5 mm} \ln\left(\frac{f \sigma_{0}^{2}}{\ell^{2}}\right) \frac{d \sigma_{0}}{d R} + \frac{8 \sigma_{0}}{3 R^{2}} \left(1 - \frac{Q^{2}}{M^{2} R}\right) + \frac{8 f}{3} \frac{d \sigma_{0}}{d R} + 2 f \ln\left(f\right) \frac{d \sigma_{0}}{d R}\right] \right. \\
& \left. + \frac{2}{R} \left[- 2 f^{-1} \hspace{0.5 mm} \ln\left(\frac{f \sigma_{0}^{2}}{\ell^{2}}\right) \frac{d \sigma_{0}}{d R} + \frac{8 \sigma_{0}}{3 R^{2}} \left(1 - \frac{Q^{2}}{M^{2} R}\right) + \frac{8 f}{3} \frac{d \sigma_{0}}{d R} + 2 f \ln\left(f\right) \frac{d \sigma_{0}}{d R}\right] \right\} \\
& \times \left\{-\frac{1}{\lambda} + \frac{f^{-2}}{32 \pi^{2}}\left(-12 \sigma_{0}^{2} \ln\left(\frac{f \sigma_{0}^{2}}{\ell^{2}}\right) + 4 \sigma_{0}^{2} + 12 \sigma_{0}^{2} \hspace{0.5 mm} f^{2} \ln f\right)\right\}^{-1} .
\end{split}
\end{align}
Below, we graph $\sigma_{0}$ and $\sigma_{2}$ for several different values of the charge-to-mass ratio $Q/M$ \cite{Mathematica}. (In all the graphs, we set $\lambda = 10^{-2}$ and $\ell = 10^{3}$.) Unlike the Schwarzschild case, the event horizon for a Reissner-Nordst\"om black hole will be located at different values of $R$ depending on the charge-to-mass ratio $Q/M$. Therefore, the horizontal axis will be in units of the dimensionless ratio $r/r_{+}$.
\begin{figure}[H]
    \centering
    \includegraphics[scale=0.62]{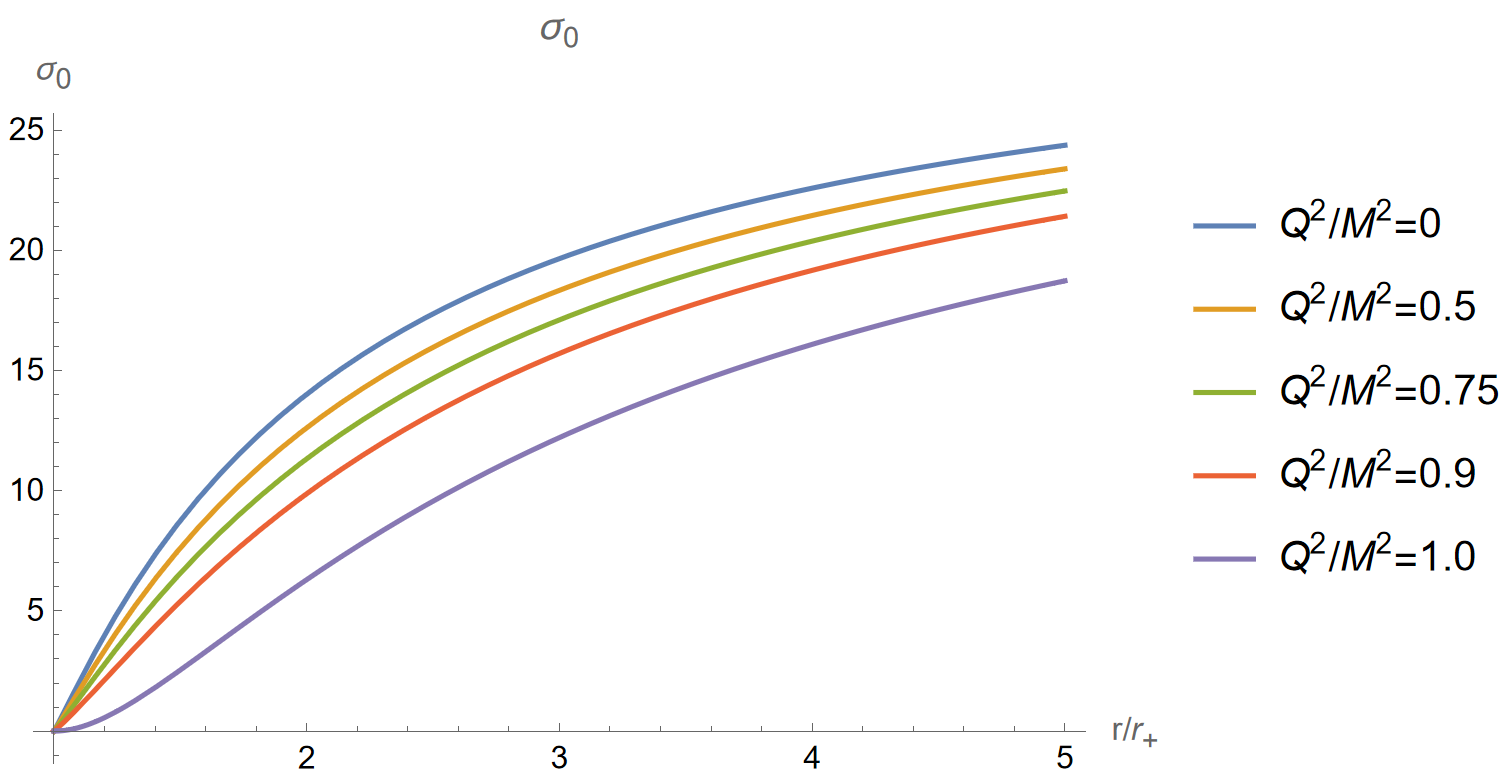}
    \caption{Graph of $\sigma_{0}$ with respect to $\frac{r}{r_{+}}$.}
    \label{fig:sigma0graphRN}
\end{figure}
\begin{figure}[H]
    \centering
    \includegraphics[scale=0.62]{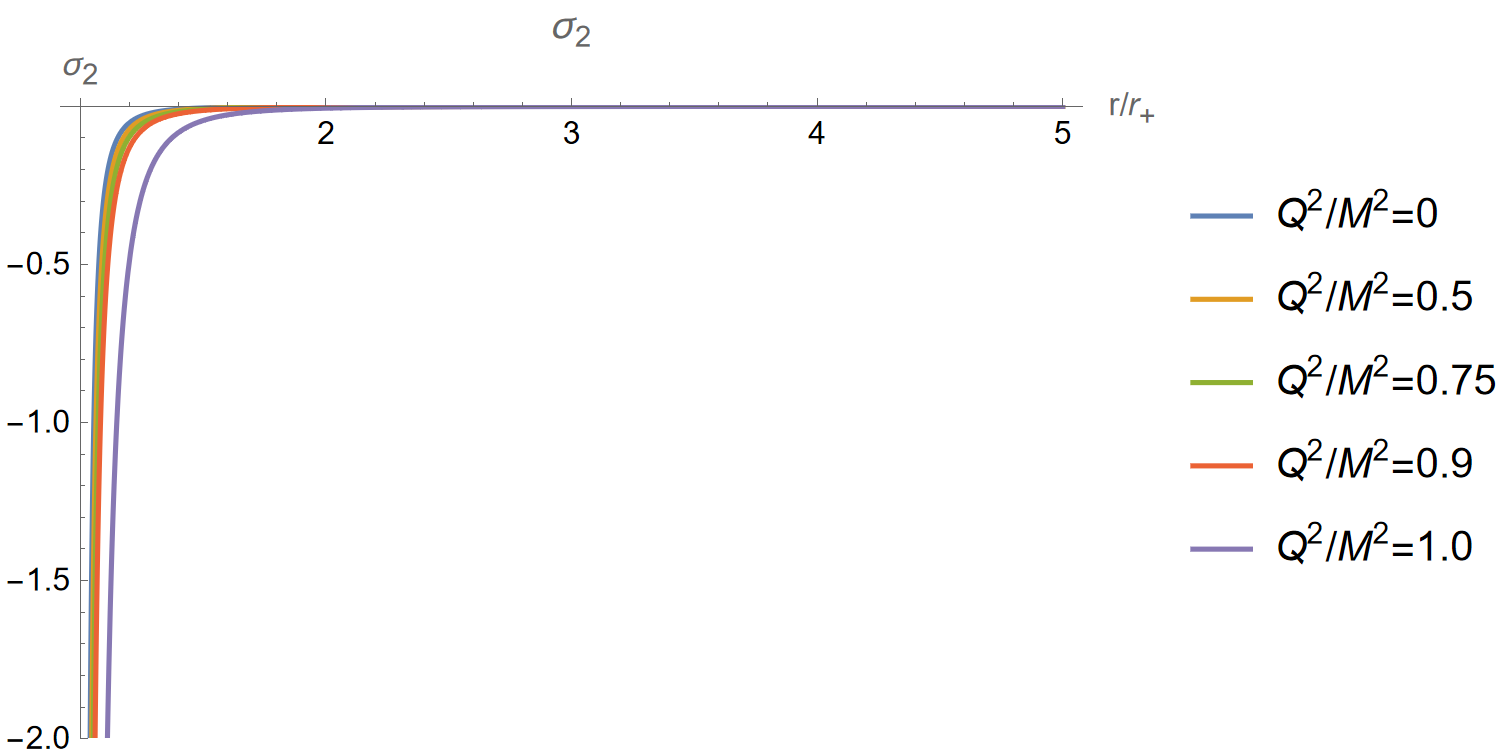}
    \caption{Graph of $\sigma_{2}$ with respect to $\frac{r}{r_{+}}$.}
    \label{fig:sigma2graphRN}
\end{figure}
\noindent In the perturbative expansion for $\sigma$, $\sigma_{2}$ has a coefficient of $M^{-2}$. Since $\sigma_{2}$ is smaller than $\sigma_{0}$ almost everywhere outside the event horizon, the profile of $\sigma$ outside the horizon is dominated by the contribution from $\sigma_{0}$. (In Appendix D, we demonstrate that the perturbative approach is valid almost everywhere outside the event horizon. Thus, higher-order effects should not significantly change the condensate profile displayed in Figure \ref{fig:sigma0graphRN}.)
\vspace{2 mm}
\newline From Figure \ref{fig:sigma0graphRN}, we see that the bubble of restored chiral symmetry becomes larger as $Q/M$ increases. This effect is particularly pronounced for near-extremal black holes ($Q/M \approx 1$), as there is a much greater difference between the $Q^{2}/M^{2} = 0.9$ and $Q^{2}/M^{2} = 1.0$ curves than between $Q^{2}/M^{2} = 0.75$ and $Q^{2}/M^{2} = 0.9$ curves.
\vspace{2 mm}
\newline For an extremal black hole, the asymptotic temperature is zero \cite{hawking1975particle, HawkingRad2}. Therefore, the local Tolman temperature is zero everywhere (see Ref. \cite{FlachiTanaka2011-chiralphasetransitions} for the definition of the Tolman temperature). However, from Figure \ref{fig:sigma0graphRN}, we see that a bubble of restored chiral symmetry still forms, surrounded by a region of spontaneously broken chiral symmetry that fills the rest of spacetime. This contradicts the usual explanation for the formation of bubbles of restored symmetry around black holes, which attributes this phenomenon to the increased local (Tolman) temperature near the event horizon.

\section{Conclusion}
In this article, we have examined the behavior of a chiral condensate outside a spherically symmetric, astrophysical-mass black hole. For a Schwarzschild black hole, the behavior of the chiral condensate is consistent with the results from Ref. \cite{FlachiTanaka2011-chiralphasetransitions}. Encouraged by this, we proceeded to analyze the behavior of a chiral condensate in the presence of a charged (Reissner-Nordstr\"om) black hole. Here, we find that radius of the bubble of restored chiral symmetry increases as the charge-to-mass ratio increases. However, the radius of the bubble never reaches infinity, even for an extremal Reissner-Nordstr\"om black hole. Thus, the chiral condensate always exhibits a phase of restored chiral symmetry close to the event horizon and a phase of spontaneously broken chiral symmetry far away from the horizon.
\vspace{2 mm}
\newline Unfortunately, it may prove difficult to test any of the predictions made in this article, since the bubble of restored chiral symmetry around an astrophysical black hole would only extend out to a few Schwarzschild radii. Therefore, in the future, we would like to investigate possible empirical signatures of chiral symmetry restoration that could be detected from Earth. For example, the restoration of chiral symmetry close to the black hole could alter the behavior of matter swirling into it, which might produce detectable radiation signatures.
\vspace{2 mm}
\newline In theory, the effect of curved spacetime on a chiral condensate would not be confined to black holes. Much subtler effects could occur due to the curvature of spacetime around the Earth. This could cause a very small difference in the proton mass (or the mass of some other hadron) depending on altitude. With sufficiently sensitive equipment, such a difference might be observable. Although these possibilities for experimental verification remain highly speculative, there is a tantalizing chance that gravity could provide an entirely new experimental window on the QCD phase diagram.

\section{Acknowledgements}
We would like to thank Dr. Laith Haddad for many interesting conversations and invaluable pieces of advice, as well as for his kind encouragement.

\printbibliography

\section*{Appendix A: Derivation of Eqn. \ref{apmeqn}}
The conformally rescaled metric is given by \cite{FlachiTanaka2011-chiralphasetransitions}
\begin{equation} \label{conformalmetric}
d \hat{s}^{2} = d t^{2} + f\left(r\right)^{-2} d r^{2} + f\left(r\right)^{-1} \hspace{0.5 mm} r^{2} d \Omega^{2} .
\end{equation}
All quantities associated with this metric will have hats. The quantity $a_{\pm}$ is defined by \cite{FlachiTanaka2011-chiralphasetransitions}
\begin{equation}
a_{\pm} = \frac{1}{180} \left(\hat{R}_{\mu \nu \tau \rho}^{2} - \hat{R}_{\mu \nu}^{2} - \hat{\Delta} \hat{R}\right) + \frac{1}{6} \hat{\Delta} \left(f \Sigma_{\pm}\right) .
\end{equation}
The Christoffel symbols are defined as (Wolfram)
\begin{equation}
\hat{\Gamma}^{m}_{i j} = \frac{1}{2} \hat{g}^{m k} \left(\frac{\partial \hat{g}_{i k}}{\partial x^{j}} + \frac{\partial \hat{g}_{j k}}{\partial x^{i}} - \frac{\partial \hat{g}_{i j}}{\partial x^{k}}\right) .
\end{equation}
Explicitly, the non-zero Christoffel symbols are given by
\begin{equation}
\hat{\Gamma}^{r}_{r r} = - \frac{f^{\prime}\left(r\right)}{f\left(r\right)} ,
\end{equation}
\begin{equation}
\hat{\Gamma}^{r}_{\theta \theta} = \frac{1}{2} r^{2} \hspace{0.5 mm} f^{\prime}\left(r\right) - r \hspace{0.5 mm} f\left(r\right) ,
\end{equation}
\begin{equation}
\hat{\Gamma}^{r}_{\phi \phi} = \sin^{2}\left(\theta\right) \left(\frac{1}{2} r^{2} 
\hspace{0.5 mm} f^{\prime}\left(r\right) - r f\left(r\right)\right) ,
\end{equation}
\begin{equation}
\hat{\Gamma}^{\theta}_{r \theta} = \Gamma^{\theta}_{\theta r} = \frac{1}{r} - \frac{f^{\prime}\left(r\right)}{2 f\left(r\right)} ,
\end{equation}
\begin{equation}
\hat{\Gamma}^{\theta}_{\phi \phi} = - \sin\left(\theta\right) \cos\left(\theta\right) ,
\end{equation}
\begin{equation}
\hat{\Gamma}^{\phi}_{r \phi} = \Gamma^{\phi}_{\phi r} = \frac{1}{r} - \frac{f^{\prime}\left(r\right)}{2 f\left(r\right)} ,
\end{equation}
\begin{equation}
\hat{\Gamma}^{\phi}_{\theta \phi} = \Gamma^{\phi}_{\phi \theta} = \cot\left(\theta\right) .
\end{equation}
The Riemann curvature tensor is defined by (Wolfram)
\begin{equation}
\hat{R}^{\alpha}_{\beta \gamma \delta} = \partial_{\gamma} \Gamma^{\alpha}_{\beta \delta} - \partial_{\delta} \Gamma^{\alpha}_{\beta \gamma} + \Gamma^{\mu}_{\beta \delta} \Gamma^{\alpha}_{\mu \gamma} - \Gamma^{\mu}_{\beta \gamma} \Gamma^{\alpha}_{\mu \delta} . 
\end{equation}
The non-zero components of the Riemann curvature tensor are given by
\begin{equation}
\hat{R}^{r}_{\theta r \theta} = -R^{r}_{\theta \theta r} = \frac{1}{2} r^{2} f^{\prime \prime}\left(r\right) - \frac{r^{2} f^{\prime}\left(r\right)^{2}}{4 f\left(r\right)} ,
\end{equation}
\begin{equation}
\hat{R}^{r}_{\phi r \phi} = - \hat{R}^{r}_{\phi \phi r} = \sin^{2}\left(\theta\right) \left(\frac{1}{2} r^{2} f^{\prime \prime}\left(r\right) - \frac{r^{2} f^{\prime}\left(r\right)^{2}}{4 f\left(r\right)}\right) ,
\end{equation}
\begin{equation}
\hat{R}^{\theta}_{r r \theta} = -\hat{R}^{\theta}_{r \theta r} = \frac{f^{\prime}\left(r\right)^{2}}{4 f\left(r\right)^{2}} - \frac{f^{\prime \prime}\left(r\right)}{2 f\left(r\right)} ,
\end{equation}
\begin{equation}
\hat{R}^{\theta}_{\phi \theta \phi} = - \hat{R}^{\theta}_{\phi \phi \theta} = \sin^{2}\left(\theta\right) \left(1 + r f^{\prime}\left(r\right) - \frac{r^{2} f^{\prime}\left(r\right)^{2}}{4 f\left(r\right)} - f\left(r\right)\right) ,
\end{equation}
\begin{equation}
\hat{R}^{\phi}_{r r \phi} = - \hat{R}^{\phi}_{r \phi r} = \frac{f^{\prime}\left(r\right)^{2}}{4 f\left(r\right)^{2}} - \frac{f^{\prime \prime}\left(r\right)}{2 f\left(r\right)} ,
\end{equation}
\begin{equation}
\hat{R}^{\phi}_{\theta \theta \phi} = - \hat{R}^{\phi}_{\theta \phi \theta} = -1 - r f^{\prime}\left(r\right) + \frac{r^{2} f^{\prime}\left(r\right)^{2}}{4 f\left(r\right)} + f\left(r\right) .
\end{equation}
The Ricci curvature tensor is defined by (Wolfram)
\begin{equation}
\hat{R}_{\mu \nu} = \hat{R}^{\lambda}_{\mu \lambda \nu} .
\end{equation}
The non-zero components of the Ricci curvature tensor are given by
\begin{equation}
\hat{R}_{r r} = - \frac{f^{\prime}\left(r\right)^{2}}{2 f\left(r\right)^{2}} + \frac{f^{\prime \prime}\left(r\right)}{f\left(r\right)} , 
\end{equation}
\begin{equation}
\hat{R}_{\theta \theta} = \frac{1}{2} r^{2} f^{\prime \prime}\left(r\right) - \frac{r^{2} f^{\prime}\left(r\right)^{2}}{2 f\left(r\right)} + r f^{\prime}\left(r\right) - f\left(r\right) + 1 ,
\end{equation}
\begin{equation}
\hat{R}_{\phi \phi} = \sin^{2}\left(\theta\right) \left(\frac{1}{2} r^{2} f^{\prime \prime}\left(r\right) - \frac{r^{2} f^{\prime}\left(r\right)^{2}}{2 f\left(r\right)} + r f^{\prime}\left(r\right) - f\left(r\right) + 1\right) .
\end{equation}
The Ricci scalar $\hat{R}$ is defined by (Wolfram)
\begin{equation}
\hat{R} = \hat{g}^{\mu \nu} \hat{R}_{\mu \nu} .
\end{equation}
Explicitly, we may write $\hat{R}$ as
\begin{equation}
\hat{R} = - \frac{3}{2} f^{\prime}\left(r\right)^{2} + 2 f\left(r\right) f^{\prime \prime}\left(r\right) + \frac{2}{r} f\left(r\right) f^{\prime}\left(r\right) - \frac{2}{r^{2}} f\left(r\right)^{2} + \frac{2}{r^{2}} f\left(r\right) .
\end{equation}
The scalars $\hat{R}^{2}_{\mu \nu \rho \tau}$ and $\hat{R}^{2}_{\mu \nu}$ are given by \cite{Mathematica}
\begin{align}
\begin{split}
\hat{R}^{2}_{\mu \nu \rho \tau} = \hspace{0.5 mm} & \frac{4 f\left(r\right)^{4}}{r^{4}} + \frac{3}{4} f^{\prime}\left(r\right)^{4} - \frac{8 f\left(r\right)^{3}}{r^{4}} - \frac{8 f\left(r\right)^{3}}{r^{3}} f^{\prime}\left(r\right) + \frac{4 f\left(r\right)^{2}}{r^{4}} + \frac{8 f\left(r\right)^{2}}{r^{3}} f^{\prime}\left(r\right) + \frac{6 f\left(r\right)^{2}}{r^{2}} f^{\prime}\left(r\right)^{2} \\
& + 2 f\left(r\right)^{2} f^{\prime \prime}\left(r\right)^{2} - \frac{2 f\left(r\right)}{r^{2}} f^{\prime}\left(r\right)^{2} - \frac{2 f\left(r\right)}{r} f^{\prime}\left(r\right)^{3} - 2 f\left(r\right) f^{\prime}\left(r\right)^{2} f^{\prime \prime}\left(r\right) ,
\end{split}
\end{align}
\begin{align}
\begin{split}
\hat{R}^{2}_{\mu \nu} = \hspace{1.0 mm} & \frac{2 f\left(r\right)^{4}}{r^{4}} + \frac{3}{4} f^{\prime}\left(r\right)^{4} - \frac{4 f\left(r\right)^{3}}{r^{4}} - \frac{4 f\left(r\right)^{3}}{r^{3}} f^{\prime}\left(r\right) - \frac{2 f\left(r\right)^{3}}{r^{2}} f^{\prime \prime}\left(r\right) + \frac{2 f\left(r\right)^{2}}{r^{4}} + \frac{4 f\left(r\right)^{2}}{r^{3}} f^{\prime}\left(r\right) \\
& - \frac{2 f\left(r\right)}{r^{2}} f^{\prime}\left(r\right)^{2} + \frac{4 f\left(r\right)^{2}}{r^{2}} f^{\prime}\left(r\right)^{2} - \frac{2 f\left(r\right)}{r} f^{\prime}\left(r\right)^{3} + \frac{2 f\left(r\right)^{2}}{r^{2}} f^{\prime \prime}\left(r\right) + \frac{2 f\left(r\right)^{2}}{r} f^{\prime}\left(r\right) f^{\prime \prime}\left(r\right) \\
& - 2 f\left(r\right) f^{\prime}\left(r\right)^{2} f^{\prime \prime}\left(r\right) + \frac{3}{2} f\left(r\right)^{2} f^{\prime \prime}\left(r\right)^{2}
\end{split}
\end{align}
The operator $\hat{\Delta}$ represents the spatial Laplace-Beltrami operator associated with the conformally rescaled metric. We only need to consider expressions in which the operator $\hat{\Delta}$ acts on scalar quantities. Thus, we may write $\hat{\Delta}$ as
\begin{equation} \label{hatDeltadef}
\hat{\Delta} = \partial_{i} \partial^{i} + \left(\partial_{i} \hat{g}^{i j}\right) \partial_{j} + \frac{\hat{g}^{i j}}{2 \det \hat{g}} \left(\partial_{i} \det \hat{g}\right) \partial_{j} ,
\end{equation}
where $i$ and $j$ are spatial indices in the conformally rescaled space-time. We only need to consider expressions in which the operator $\hat{\Delta}$ acts on scalar functions that depend only on $r$. Thus, we may rewrite Eqn. \ref{hatDeltadef} as
\begin{equation} \label{hatDelta2}
\hat{\Delta} = g^{r r} \partial^{2}_{r} + \left(\partial_{i} \hat{g}^{i r}\right) \partial_{r} + \frac{\hat{g}^{i r}}{2 \det \hat{g}} \left(\partial_{i} \det \hat{g}\right) \partial_{r} .
\end{equation}
Plugging Eqn. \ref{conformalmetric} into Eqn. \ref{hatDelta2}, we find that
\begin{equation}
\hat{\Delta} = f\left(r\right)^{2} \frac{d^{2}}{d r^{2}} + \frac{2 f\left(r\right)^{2}}{r} \frac{d}{d r} = f\left(r\right)^{2} \Delta_{\textrm{Eucl}} ,
\end{equation}
where $\Delta_{\textrm{Eucl}}$ is the Laplacian for three-dimensional Euclidean space. Finally, we may write $a_{\pm}$ as
\begin{align}
\begin{split}
a_{\pm} = \hspace{1.0 mm} & \frac{1}{180} \left\{ \frac{2 f\left(r\right)^{2}}{r^{4}} - \frac{8 f\left(r\right)^{3}}{r^{4}} + \frac{6 f\left(r\right)^{4}}{r^{4}} + \frac{8 f\left(r\right)^{2}}{r^{3}} f^{\prime}\left(r\right) - \frac{12 f\left(r\right)^{3}}{r^{3}} f^{\prime}\left(r\right) + \frac{6 f\left(r\right)^{2}}{r^{2}} f^{\prime}\left(r\right)^{2} - \frac{4 f\left(r\right)^{2}}{r^{2}} f^{\prime \prime}\left(r\right) \right. \\
& \left. + \frac{6 f\left(r\right)^{3}}{r^{2}} f^{\prime \prime}\left(r\right) - \frac{6 f\left(r\right)^{2}}{r} f^{\prime}\left(r\right) f^{\prime \prime}\left(r\right) + \frac{3}{2} f\left(r\right)^{2} f^{\prime \prime}\left(r\right)^{2} - \frac{6 f\left(r\right)^{3}}{r} f^{\left(3\right)}\left(r\right) - f\left(r\right)^{2} f^{\prime}\left(r\right) f^{\left(3\right)}\left(r\right) \right. \\
& \left. - 2 f\left(r\right)^{3} f^{\left(4\right)}\left(r\right) \right\} + \frac{f\left(r\right)^{2}}{6} \Delta_{\textrm{Eucl}}\left(f \Sigma_{\pm}\right) .
\end{split}
\end{align}

\section*{Appendix B: Computation of $\sigma_{0}\left(r\right)$}
At zeroth order in $M^{-2}$, we may write the Euler-Lagrange equations as
\begin{equation} \label{EL0order}
- \frac{1}{2 \lambda} + \frac{f^{-2}}{16 \pi^{2}} \left(\sigma_{0}^{2} - \sigma_{0}^{2} \ln\left(\frac{f \sigma_{0}^{2}}{\ell^{2}}\right) + f^{2} \ln\left(f\right) \sigma_{0}^{2}\right) = 0 .
\end{equation}
Through simple algebra, we may rewrite Eqn. \ref{EL0order} as
\begin{equation} \label{EL01}
\frac{8 \pi^{2} f^{2}}{\lambda \sigma_{0}^{2}} + \ln\left(\sigma_{0}^{2}\right) = 1 - \ln\left(\frac{f}{\ell^{2}}\right) + f^{2} \ln\left(f\right) .
\end{equation}
In Eqn. \ref{EL01}, we would like the argument of the logarithm on the LHS to be equal to the first term on the LHS. With a bit more algebra, we may rewrite Eqn. \ref{EL01} as
\begin{equation} \label{EL02}
-\frac{8 \pi^{2} f^{2}}{\lambda \sigma_{0}^{2}} + \ln\left(-\frac{8 \pi^{2} f^{2}}{\lambda \sigma_{0}^{2}}\right) = -1 + \ln\left(\frac{f}{\ell^{2}}\right) - f^{2} \ln\left(f\right) + \ln\left(-\frac{8 \pi^{2} f^{2}}{\lambda}\right) .
\end{equation}
Next, we take the exponential of both sides of Eqn. \ref{EL02} to obtain
\begin{equation} \label{EL03}
-\frac{8 \pi^{2} f^{2}}{\lambda \sigma_{0}^{2}} \exp\left(-\frac{8 \pi^{2} f^{2}}{\lambda \sigma_{0}^{2}}\right) = -\frac{8 \pi^{2}}{e \hspace{0.5 mm} \lambda \hspace{0.5 mm} \ell^{2}} f^{3 - f^{2}} .
\end{equation}
The Lambert W function is defined by \cite{Wfuncref}
\begin{equation}
w\left(x\right) \exp\left(w\left(x\right)\right) = x .
\end{equation}
Thus, we may rewrite Eqn. \ref{EL03} as
\begin{equation}
-\frac{8 \pi^{2} f^{2}}{\lambda \sigma_{0}^{2}} = w\left(-\frac{8 \pi^{2}}{e  \lambda \ell^{2}} f^{3 - f^{2}}\right) ,
\end{equation}
or equivalently,
\begin{equation} \label{sigmaminexplicit}
\sigma_{0}^{2} = -\frac{8 \pi^{2} f^{2}}{\lambda} \left[w\left(-\frac{8 \pi^{2}}{e \lambda \ell^{2}} f^{3 - f^{2}}\right)\right]^{-1} .
\end{equation}
Because we are working in Euclideanized space-time, $f$ is positive everywhere. Therefore, the argument of $w\left(x\right)$ in Eqn. \ref{sigmaminexplicit} will be negative everywhere. For negative arguments, the Lambert W function has two branches: $W_{0}\left(x\right)$ and $W_{-1}\left(x\right)$. From previous numerical results, we expect that there exists a region of restored chiral symmetry close to the event horizon \cite{FlachiTanaka2011-chiralphasetransitions}. Below, we graph both solutions for $\sigma_{0}^{2}\left(R\right)$ (corresponding to both branches of the Lambert W function) \cite{Mathematica}.
\begin{figure}[H]
    \centering
    \includegraphics[scale=0.46]{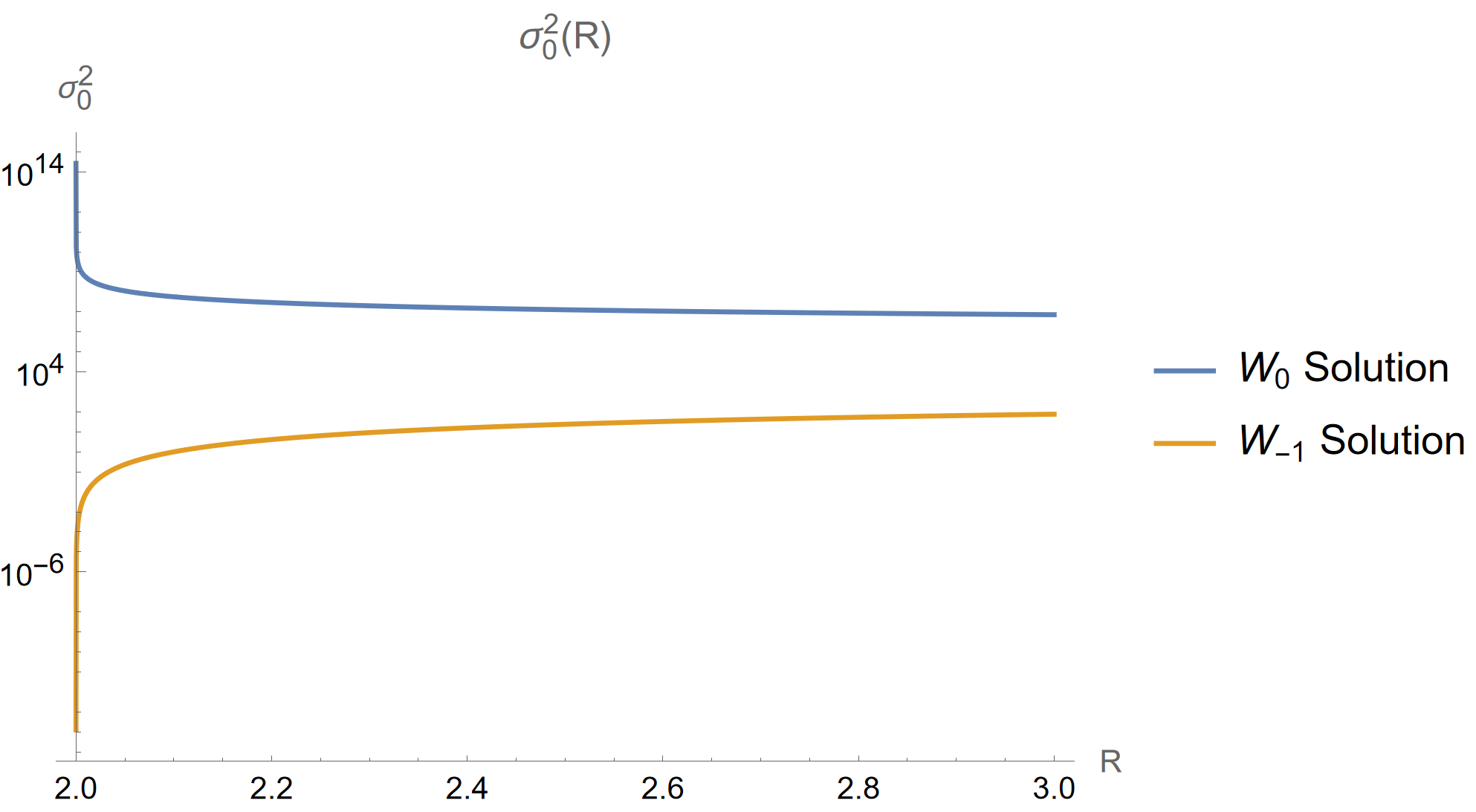}
    \caption{Comparison of the two possible solutions for $\sigma_{0}^{2}\left(R\right)$}
    \label{fig:Sigma0comp}
\end{figure}
\noindent From Figure \ref{fig:Sigma0comp}, we see that only the $W_{-1}$ solution restores chiral symmetry close to the event horizon. Thus, we may rewrite Eqn. \ref{sigmaminexplicit} as
\begin{equation}
\sigma_{0}^{2} = -\frac{8 \pi^{2} f^{2}}{\lambda} \left[W_{-1}\left(-\frac{8 \pi^{2}}{e \lambda \ell^{2}} f^{3 - f^{2}}\right)\right]^{-1} .
\end{equation}

\section*{Appendix C: Derivation of Eqns. \ref{sigma0solRN} and \ref{sigma2solRN}}
In the original coordinate system (with the original radial coordinate $r$), the quantity $a_{\textrm{sp}}$ takes the form \cite{Mathematica}
\begin{equation}
a_{\textrm{sp}} = \frac{2 f\left(r\right)^{2}}{5 r^{8}} \left[\left(2 M^{2} - Q^{2}\right) r^{2} - 2 M Q^{2} r + Q^{4}\right] .
\end{equation}
In the new coordinate system (with the rescaled radial coordinate $R$), the quantity $a_{\textrm{sp}}$ takes the form
\begin{equation}
a_{\textrm{sp}} = \frac{2 f\left(r\right)^{2}}{5 M^{8} R^{8}} \left[\left(2 M^{2} - Q^{2}\right) M^{2} R^{2} - 2 M^{2} Q^{2} R + Q^{4}\right] .
\end{equation}
Since $Q$ can be no larger than $M$, $a_{\textrm{sp}}$ must be of order $M^{-4}$ or smaller. We will approximate the Lagrangian to order $M^{-2}$. Therefore, we will neglect the contribution of $a_{\textrm{sp}}$ to the Lagrangian. Keeping terms up to $M^{-2}$, we may write the Lagrangian pieces $\mathcal{A}$, $\mathcal{B}$, and $\delta \mathcal{L}$ as \cite{Mathematica}
\begin{equation} \label{ARN}
\mathcal{A} = \left(\frac{3}{2} - \ln\left(\frac{f}{\ell^{2}}\right)\right) \left(\sigma^{4} + \frac{f}{M^{2}} \left(\frac{d \sigma}{d R}\right)^{2}\right) - \frac{1}{2} \left(\Sigma_{+}^{2} \ln \Sigma_{+} + \Sigma_{-}^{2} \ln \Sigma_{-}\right) ,
\end{equation}
\begin{equation} \label{BRN}
\mathcal{B} = \frac{8 \sigma}{3 M^{2} R^{2}} \left(1 - \frac{Q^{2}}{M^{2} R}\right) \frac{d \sigma}{d R} + \frac{4 \sigma^{2}}{3 M^{2} \hspace{0.5 mm} R^{4} f} \left(1 - \frac{Q^{2}}{M^{2} R}\right)^{2} + \frac{f}{6 M^{2} \hspace{0.5 mm} \Sigma_{+}} \left(\frac{d \Sigma_{+}}{d R}\right)^{2} + \frac{f}{6 M^{2} \hspace{0.5 mm} \Sigma_{-}} \left(\frac{d \Sigma_{-}}{d R}\right)^{2} ,
\end{equation}
\begin{equation} \label{dLRN}
\delta \mathcal{L} = \left(\sigma^{4} + \frac{f}{M^{2}} \left(\frac{d \sigma}{d R}\right)^{2}\right) \ln f - \frac{\sigma^{2}}{3 M^{2} \hspace{0.5 mm} f} \left(\frac{8}{R^{3}} - \frac{14}{R^{4}} + \frac{24 Q^{2}}{M^{2} R^{5}} - \frac{10 Q^{2}}{M^{2} R^{4}} - \frac{8 Q^{4}}{M^{4} R^{6}}\right) .
\end{equation}
In Eqns. \ref{ARN}-\ref{dLRN}, we have accounted for the fact that $Q$ and $M$ are at similar scales for a near-extremal RN black hole. Even though we have approximated the Lagrangian to the order of $M^{-2}$, some terms contain factors of $M^{-4}$ or $M^{-6}$. This occurs because of the possibility that factors of $Q$ could ``raise" these terms up to order $M^{-2}$ or higher. As in the Schwarzschild case, we may write $\Sigma_{\pm}$ as
\begin{equation}
\Sigma_{\pm} = \sigma^{2} \left(1 \pm \frac{\sqrt{f}}{M \sigma^{2}} \frac{d \sigma}{d R}\right) .
\end{equation}
In Eqns. \ref{lnSigmataylor} and \ref{Sigmainvtaylor}, we derived Taylor series approximations for $\ln \Sigma_{\pm}$ and $\Sigma_{\pm}^{-1}$. Fortunately, these approximations are still valid in RN spacetime. Explicitly,
\begin{equation}
\ln \Sigma_{\pm} = \ln\left(\sigma^{2}\right) \pm \frac{\sqrt{f}}{M \sigma^{2}} \frac{d \sigma}{d R} - \frac{f}{2 M^{2} \sigma^{4}} \left(\frac{d \sigma}{d R}\right)^{2} ,
\end{equation}
\begin{equation}
\Sigma_{\pm}^{-1} = \frac{1}{\sigma^{2}} \left(1 \mp \frac{\sqrt{f}}{M \sigma^{2}} \frac{d \sigma}{d R} + \frac{f}{M^{2} \sigma^{4}} \left(\frac{d \sigma}{d R}\right)^{2}\right) .
\end{equation}
Thus, we may rewrite Eqns. \ref{ARN}-\ref{dLRN} as
\begin{equation}
\mathcal{A} = - \ln\left(\frac{f \sigma^{2}}{\ell^{2}}\right) \left(\sigma^{4} + \frac{f}{M^{2}} \left(\frac{d \sigma}{d R}\right)^{2}\right) + \frac{3}{2} \sigma^{4} ,
\end{equation}
\begin{equation}
\mathcal{B} = \frac{8 \sigma}{3 M^{2} R^{2}} \left(1 - \frac{Q^{2}}{M^{2} R}\right) \frac{d \sigma}{d R} + \frac{4 \sigma^{2}}{3 M^{2} \hspace{0.5 mm} R^{4} f} \left(1 - \frac{Q^{2}}{M^{2} R}\right)^{2} + \frac{4 f}{3 M^{2}} \left(\frac{d \sigma}{d R}\right)^{2} ,
\end{equation}
\begin{equation}
\delta \mathcal{L} = \left(\sigma^{4} + \frac{f}{M^{2}} \left(\frac{d \sigma}{d R}\right)^{2}\right) \ln f - \frac{\sigma^{2}}{3 M^{2} \hspace{0.5 mm} f} \left(\frac{8}{R^{3}} - \frac{14}{R^{4}} + \frac{24 Q^{2}}{M^{2} R^{5}} - \frac{10 Q^{2}}{M^{2} R^{4}} - \frac{8 Q^{4}}{M^{4} R^{6}}\right) .
\end{equation}
Using Eqn. \ref{Lschwarznew}, we may write the Lagrangian as
\begin{align} \label{LRN}
\begin{split}
\mathcal{L} = & -\frac{\sigma^{2}}{2 \lambda} + \frac{f^{-2}}{32 \pi^{2}} \left(- \ln\left(\frac{f \sigma^{2}}{\ell^{2}}\right) \left(\sigma^{4} + \frac{f}{M^{2}} \left(\frac{d \sigma}{d R}\right)^{2}\right) + \frac{3}{2} \sigma^{4}\right) \\
& + \frac{1}{32 \pi^{2}} \left(\frac{8 \sigma}{3 M^{2} R^{2}} \left(1 - \frac{Q^{2}}{M^{2} R}\right) \frac{d \sigma}{d R} + \frac{4 \sigma^{2}}{3 M^{2} \hspace{0.5 mm} R^{4} f} \left(1 - \frac{Q^{2}}{M^{2} R}\right)^{2} + \frac{4 f}{3 M^{2}} \left(\frac{d \sigma}{d R}\right)^{2}\right) \\
& + \frac{1}{32 \pi^{2}} \left(\left(\sigma^{4} + \frac{f}{M^{2}} \left(\frac{d \sigma}{d R}\right)^{2}\right) \ln f - \frac{\sigma^{2}}{3 M^{2} \hspace{0.5 mm} f} \left(\frac{8}{R^{3}} - \frac{14}{R^{4}} + \frac{24 Q^{2}}{M^{2} R^{5}} - \frac{10 Q^{2}}{M^{2} R^{4}} - \frac{8 Q^{4}}{M^{4} R^{6}}\right)\right) .
\end{split}
\end{align}
As in the Schwarzschild case, the Euler-Lagrange equation of motion is given by
\begin{equation} \label{EOMRN}
\frac{\partial \mathcal{L}}{\partial \sigma} - \frac{d}{d R} \left(\frac{\partial \mathcal{L}}{\partial \sigma^{\prime}}\right) - \frac{2}{R} \frac{\partial \mathcal{L}}{\partial \sigma^{\prime}} = 0 .
\end{equation}
Plugging Eqn. \ref{LRN} into Eqn. \ref{EOMRN}, we find that
\begin{align}
\begin{split}
& - \frac{\sigma}{\lambda} + \frac{f^{-2}}{32 \pi^{2}} \left(- 4 \sigma^{3} \ln\left(\frac{f \sigma^{2}}{\ell^{2}}\right) + 4 \sigma^{3} - \frac{2 f}{M^{2} \hspace{0.5 mm} \sigma} \left(\frac{d \sigma}{d R}\right)^{2} + \frac{8 f^{2}}{3 M^{2} R^{2}} \left(1 - \frac{Q^{2}}{M^{2} R}\right) \frac{d \sigma}{d R}\right) \\
& + \frac{1}{32 \pi^{2}} \left(\frac{8 \sigma}{3 M^{2} \hspace{0.5 mm} R^{4} f} \left(1 - \frac{Q^{2}}{M^{2} R}\right)^{2} + 4 \sigma^{3} \ln f - \frac{2 \sigma}{3 M^{2} \hspace{0.5 mm} f} \left(\frac{8}{R^{3}} - \frac{14}{R^{4}} + \frac{24 Q^{2}}{M^{2} R^{5}} - \frac{10 Q^{2}}{M^{2} R^{4}} - \frac{8 Q^{4}}{M^{4} R^{6}}\right)\right) \\
& - \frac{1}{32 \pi^{2}} \frac{d}{d R} \left\{- \ln\left(\frac{f \sigma^{2}}{\ell^{2}}\right) \frac{2 f^{-1}}{M^{2}} \frac{d \sigma}{d R} + \frac{8 \sigma}{3 M^{2} R^{2}} \left(1 - \frac{Q^{2}}{M^{2} R}\right) + \frac{8 f}{3 M^{2}} \frac{d \sigma}{d R} + \frac{2 f \ln f}{M^{2}} \frac{d \sigma}{d R}\right\} \\
& - \frac{1}{16 \pi^{2} R} \left\{- \ln\left(\frac{f \sigma^{2}}{\ell^{2}}\right) \frac{2 f^{-1}}{M^{2}} \frac{d \sigma}{d R} + \frac{8 \sigma}{3 M^{2} R^{2}} \left(1 - \frac{Q^{2}}{M^{2} R}\right) + \frac{8 f}{3 M^{2}} \frac{d \sigma}{d R} + \frac{2 f \ln f}{M^{2}} \frac{d \sigma}{d R}\right\} = 0 .
\end{split}
\end{align}
At zeroth order in $M^{-1}$, we may write the equation of motion as
\begin{equation} \label{EOM01RN}
- \frac{\sigma_{0}}{\lambda} + \frac{f^{-2}}{32 \pi^{2}} \left(-4 \sigma_{0}^{3} \ln\left(\frac{f \sigma_{0}^{2}}{\ell^{2}}\right) + 4 \sigma_{0}^{3}\right) + \frac{1}{32 \pi^{2}} \left(4 \sigma_{0}^{3} \ln f\right) = 0 .
\end{equation}
We look for non-trival solutions, where $\sigma$ is not constant and approaches a non-zero value at spatial infinity. Therefore, we may rewrite Eqn. \ref{EOM01RN} as
\begin{equation} \label{EOM02RN}
- \frac{1}{2 \lambda} + \frac{f^{-2}}{16 \pi^{2}} \left(\sigma_{0}^{2} - \sigma_{0}^{2} \ln\left(\frac{f \sigma_{0}^{2}}{\ell^{2}}\right) + f^{2} \ln\left(f\right) \sigma_{0}^{2}\right) = 0 .
\end{equation}
It is easy to see that Eqn. \ref{EOM02RN} is equivalent to Eqn. \ref{EOM02}, except that $f\left(r\right)$ is now given by the Reissner-Nordstr\"om metric instead of the Schwarzschild metric. Therefore, we may write $\sigma_{0}$ as
\begin{equation}
\sigma_{0} = \sqrt{-\frac{8 \pi^{2} f^{2}}{\lambda} \left[W_{-1}\left(-\frac{8 \pi^{2}}{e \lambda \ell^{2}} f^{3 - f^{2}}\right)\right]^{-1}} .
\end{equation}
Using the Mercator series, we may write $\ln \sigma$ as
\begin{equation}
\ln \sigma = \ln \sigma_{0} + \frac{\sigma_{2}\left(r\right)}{M^{2} \hspace{0.5 mm} \sigma_{0}\left(r\right)} - \frac{1}{2} \left(\frac{\sigma_{2}\left(r\right)}{M^{2} \hspace{0.5 mm} \sigma_{0}\left(r\right)}\right)^{2} + \dots
\end{equation}
At second order in $M^{-1}$, we may write the equation of motion as
\begin{align}
\begin{split}
& - \frac{\sigma_{2}}{\lambda} + \frac{f^{-2}}{32 \pi^{2}} \left(-12 \sigma_{0}^{2} \sigma_{2} \ln\left(\frac{f \sigma_{0}^{2}}{\ell^{2}}\right) + 4 \sigma_{0}^{2} \sigma_{2} - \frac{2 f}{\sigma_{0}} \left(\frac{d \sigma_{0}}{d R}\right)^{2} + \frac{8 f^{2}}{3 R^{2}} \left(1 - \frac{Q^{2}}{M^{2} R}\right) \frac{d \sigma_{0}}{d R}\right) \\
& + \frac{1}{32 \pi^{2}} \left(\frac{8 \sigma_{0}}{3 R^{4} f} \left(1 - \frac{Q^{2}}{M^{2} R}\right)^{2} + 12 \sigma_{0}^{2} \sigma_{2} \ln f - \frac{2 \sigma_{0}}{3 f} \left(\frac{8}{R^{3}} - \frac{14}{R^{4}} + \frac{24 Q^{2}}{M^{2} R^{5}} - \frac{10 Q^{2}}{M^{2} R^{4}} - \frac{8 Q^{4}}{M^{4} R^{6}}\right)\right) \\
& - \frac{1}{32 \pi^{2}} \frac{d}{d R} \left\{- 2 f^{-1} \hspace{0.5 mm} \ln\left(\frac{f \sigma_{0}^{2}}{\ell^{2}}\right) \frac{d \sigma_{0}}{d R} + \frac{8 \sigma_{0}}{3 R^{2}} \left(1 - \frac{Q^{2}}{M^{2} R}\right) + \frac{8 f}{3} \frac{d \sigma_{0}}{d R} + 2 f \ln\left(f\right) \frac{d \sigma_{0}}{d R}\right\} \\
& - \frac{1}{16 \pi^{2} R} \left\{- 2 f^{-1} \hspace{0.5 mm} \ln\left(\frac{f \sigma_{0}^{2}}{\ell^{2}}\right) \frac{d \sigma_{0}}{d R} + \frac{8 \sigma_{0}}{3 R^{2}} \left(1 - \frac{Q^{2}}{M^{2} R}\right) + \frac{8 f}{3} \frac{d \sigma_{0}}{d R} + 2 f \ln\left(f\right) \frac{d \sigma_{0}}{d R}\right\} = 0 .
\end{split}
\end{align}
Using simple algebra, we may write the solution $\sigma_{2}$ as
\begin{align}
\begin{split}
\sigma_{2} = & \frac{1}{32 \pi^{2}}\left\{\frac{2 f^{-1}}{\sigma_{0}} \left(\frac{d \sigma_{0}}{d R}\right)^{2} - \frac{8}{3 R^{2}} \left(1 - \frac{Q^{2}}{M^{2} R}\right) \frac{d \sigma_{0}}{d R} \right. \\
& \left. -\frac{8 \sigma_{0}}{3 R^{4} f} \left(1 - \frac{Q^{2}}{M^{2} R}\right)^{2} + \frac{2 \sigma_{0}}{3 f} \left(\frac{8}{R^{3}} - \frac{14}{R^{4}} + \frac{24 Q^{2}}{M^{2} R^{5}} - \frac{10 Q^{2}}{M^{2} R^{4}} - \frac{8 Q^{4}}{M^{4} R^{6}}\right) \right. \\
& \left. + \frac{d}{d R} \left[- 2 f^{-1} \hspace{0.5 mm} \ln\left(\frac{f \sigma_{0}^{2}}{\ell^{2}}\right) \frac{d \sigma_{0}}{d R} + \frac{8 \sigma_{0}}{3 R^{2}} \left(1 - \frac{Q^{2}}{M^{2} R}\right) + \frac{8 f}{3} \frac{d \sigma_{0}}{d R} + 2 f \ln\left(f\right) \frac{d \sigma_{0}}{d R}\right] \right. \\
& \left. + \frac{2}{R} \left[- 2 f^{-1} \hspace{0.5 mm} \ln\left(\frac{f \sigma_{0}^{2}}{\ell^{2}}\right) \frac{d \sigma_{0}}{d R} + \frac{8 \sigma_{0}}{3 R^{2}} \left(1 - \frac{Q^{2}}{M^{2} R}\right) + \frac{8 f}{3} \frac{d \sigma_{0}}{d R} + 2 f \ln\left(f\right) \frac{d \sigma_{0}}{d R}\right] \right\} \\
& \times \left\{-\frac{1}{\lambda} + \frac{f^{-2}}{32 \pi^{2}}\left(-12 \sigma_{0}^{2} \ln\left(\frac{f \sigma_{0}^{2}}{\ell^{2}}\right) + 4 \sigma_{0}^{2} + 12 \sigma_{0}^{2} \hspace{0.5 mm} f^{2} \ln f\right)\right\}^{-1} .
\end{split}
\end{align}

\section*{Appendix D: Verification of the Perturbative Approach for a Large Reissner-Nordstr\"om Black Hole}
To obtain the results of Appendix C, we had to make three assumptions. First, we assumed that $\beta \sqrt{f \Sigma_{\pm}} \gg 1$ for both $\Sigma_{+}$ and $\Sigma_{-}$. Second, we assumed that $\sigma$ varies on scales far larger than the Planck scale. Lastly, we assumed that $\sigma$ can be expanded perturbatively in powers of $M^{-2}$. We wish to verify that these assumptions hold almost everywhere outside the event horizon. 
\vspace{2 mm}
\newline Let us first verify the assumption $\beta \sqrt{f \Sigma_{\pm}} \gg 1$. As discussed earlier, an astrophysical black hole should have a mass of at least $\sim 10^{38}$ Planck masses. The inverse temperature $\beta$ is given by \cite{hawking1975particle, HawkingRad2}
\begin{equation} \label{betaRNeqn}
\beta = 4 \pi r_{+} \hspace{0.5 mm} \left(1 - \frac{Q^{2}}{r_{+}^{2}}\right)^{-1} .
\end{equation}
From Eqns. \ref{betaRNeqn} and \ref{RNEHrad}, we see that $\beta \geq 4 \pi M$. Therefore, $\beta$ should be at least $\sim 10^{39}$. Hence, we see that the assumption $\beta \sqrt{f \Sigma_{\pm}} \gg 1$ is satisfied if $f \Sigma_{\pm} \gg 10^{-78}$. We may write $\Sigma_{\pm}$ as
\begin{equation}
\Sigma_{\pm} = \sigma^{2} \pm \frac{\sqrt{f}}{M} \frac{d \sigma}{d R} .
\end{equation}
Next, we use Eqn. \ref{sigmaexpansion} to expand $\Sigma_{\pm}$ in powers of $M^{-1}$. Keeping terms up to $M^{-2}$, we find that
\begin{equation}
\Sigma_{\pm} = \sigma_{0}^{2} + \frac{2}{M^{2}} \sigma_{0} \sigma_{2} \pm \frac{\sqrt{f}}{M} \frac{d \sigma_{0}}{d R} .
\end{equation}
Fortunately, we already have expressions for $\sigma_{0}$ and $\sigma_{2}$. Below, we graph the functions $f\left(R\right) \Sigma_{+}\left(R\right)$ and $f\left(R\right) \Sigma_{-}\left(R\right)$ for $M = 10^{39}$ Planck masses and several different values of the charge ratio $Q/M$ \cite{Mathematica}.
\begin{figure}[H]
    \centering
    \includegraphics[scale=0.6]{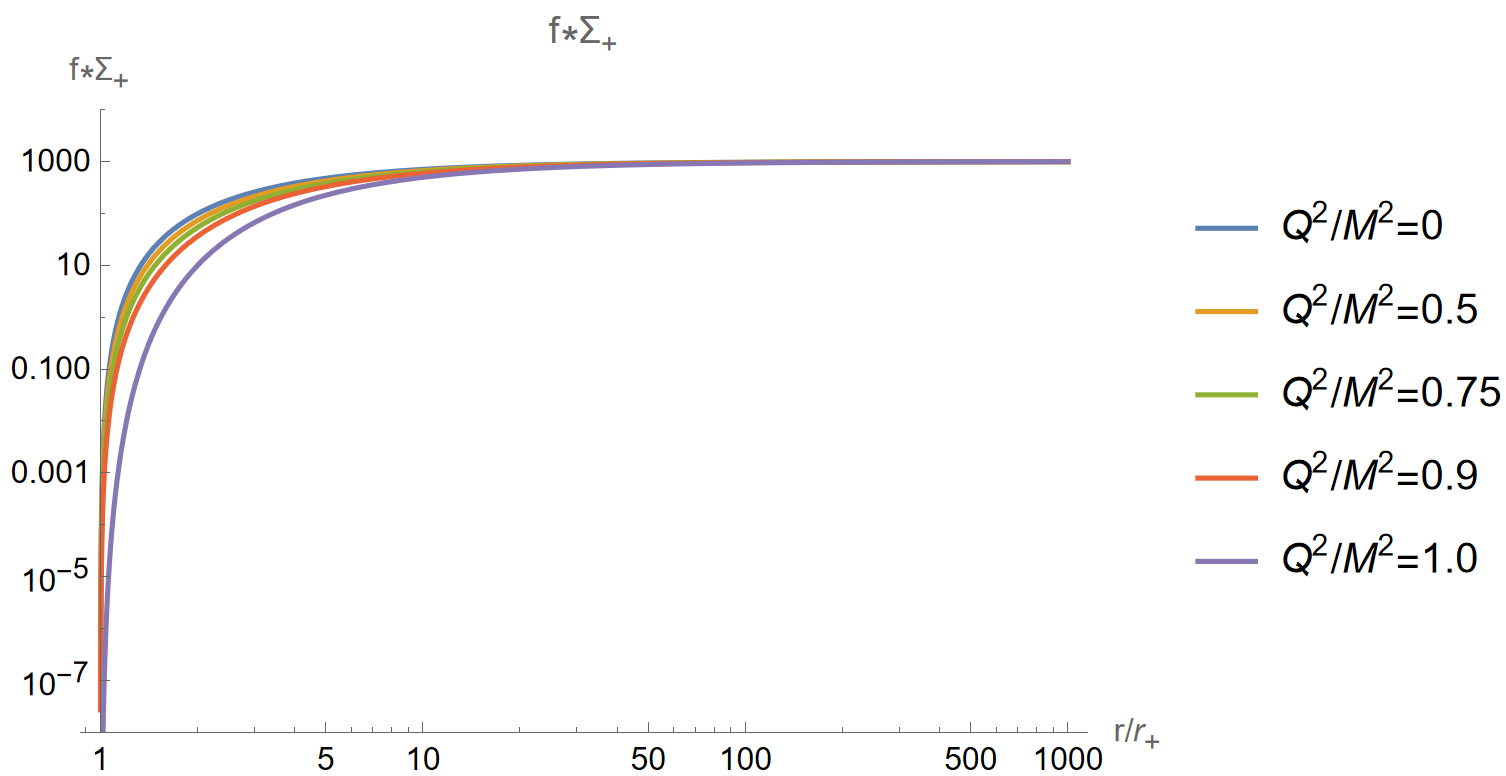}
    \caption{Graph of $f * \Sigma_{+}$ with respect to $r/r_{+}$, with $\lambda = 10^{-2}$ and $\ell = 10^{3}$.}
    \label{fig:SigmaPlusRN}
\end{figure}
\begin{figure}[H]
    \centering
    \includegraphics[scale=0.6]{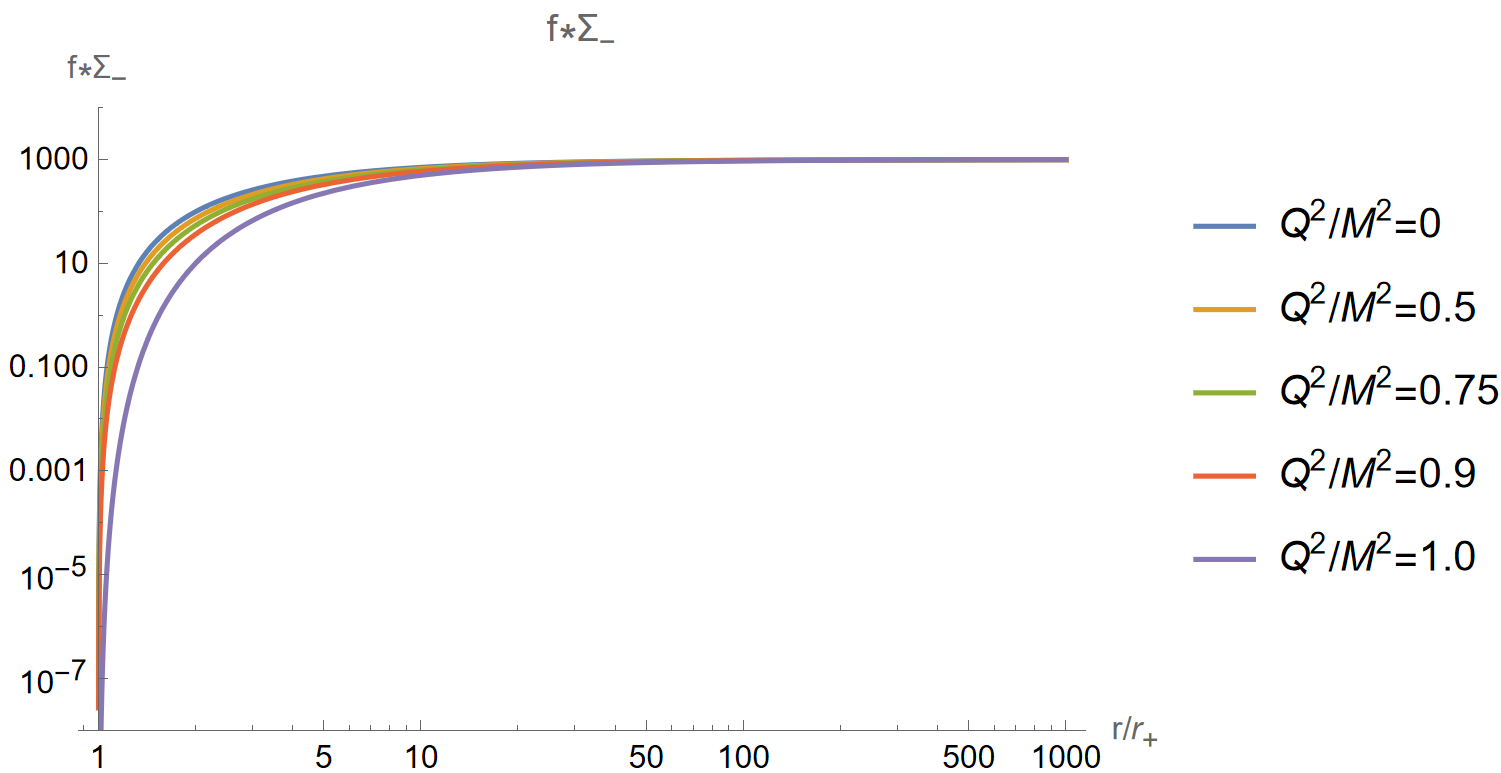}
    \caption{Graph of $f * \Sigma_{-}$ with respect to $r/r_{+}$, with $\lambda = 10^{-2}$ and $\ell = 10^{3}$.}
    \label{fig:SigmaMinusRN}
\end{figure}
\noindent From Figures \ref{fig:SigmaPlusGraph} and \ref{fig:SigmaMinusGraph}, we see that the assumption $\beta \sqrt{f \Sigma_{\pm}} \gg 1$ may be violated very close to the event horizon. Nevertheless, it is also clear from these graphs that the assumption $\beta \sqrt{f \Sigma_{\pm}} \gg 1$ holds almost everywhere outside the event horizon.
\vspace{2 mm}
\newline Next, let us verify the assumption that $\sigma$ varies on scales far larger than the Planck scale. In terms of the original Planck-scale coordinate $r$, we may write this condition as
\begin{equation} \label{AppDdsdrll1}
\frac{d \sigma}{d r} \ll 1 .
\end{equation}
Rewriting Eqn. \ref{AppDdsdrll1} in terms of the rescaled coordinate $R$, we obtain
\begin{equation} \label{AppDdsdRll1}
\frac{d \sigma}{d R} \ll M .
\end{equation}
Up to second-order in $M^{-1}$, we may write $\sigma$ as
\begin{equation}
\sigma\left(R\right) = \sigma_{0}\left(R\right) + \frac{\sigma_{2}\left(R\right)}{M^{2}} .
\end{equation}
Fortunately, we already have expressions for $\sigma_{0}$ and $\sigma_{2}$. Below, we graph $\sigma^{\prime}\left(R\right)$ for $M = 10^{39}$ Planck masses and several different values of $Q/M$ \cite{Mathematica}. From this graph, we see that Eqn. \ref{AppDdsdRll1} is satisfied everywhere outside the event horizon. (To account for the different event horizon radii for different values of $Q/M$, we will plot all the curves with respect to $r/r_{+}$. However, the curves will still represent the derivative of $\sigma\left(R\right)$ with respect to $R$.) 
\begin{figure}[H]
    \centering
    \includegraphics[scale=0.62]{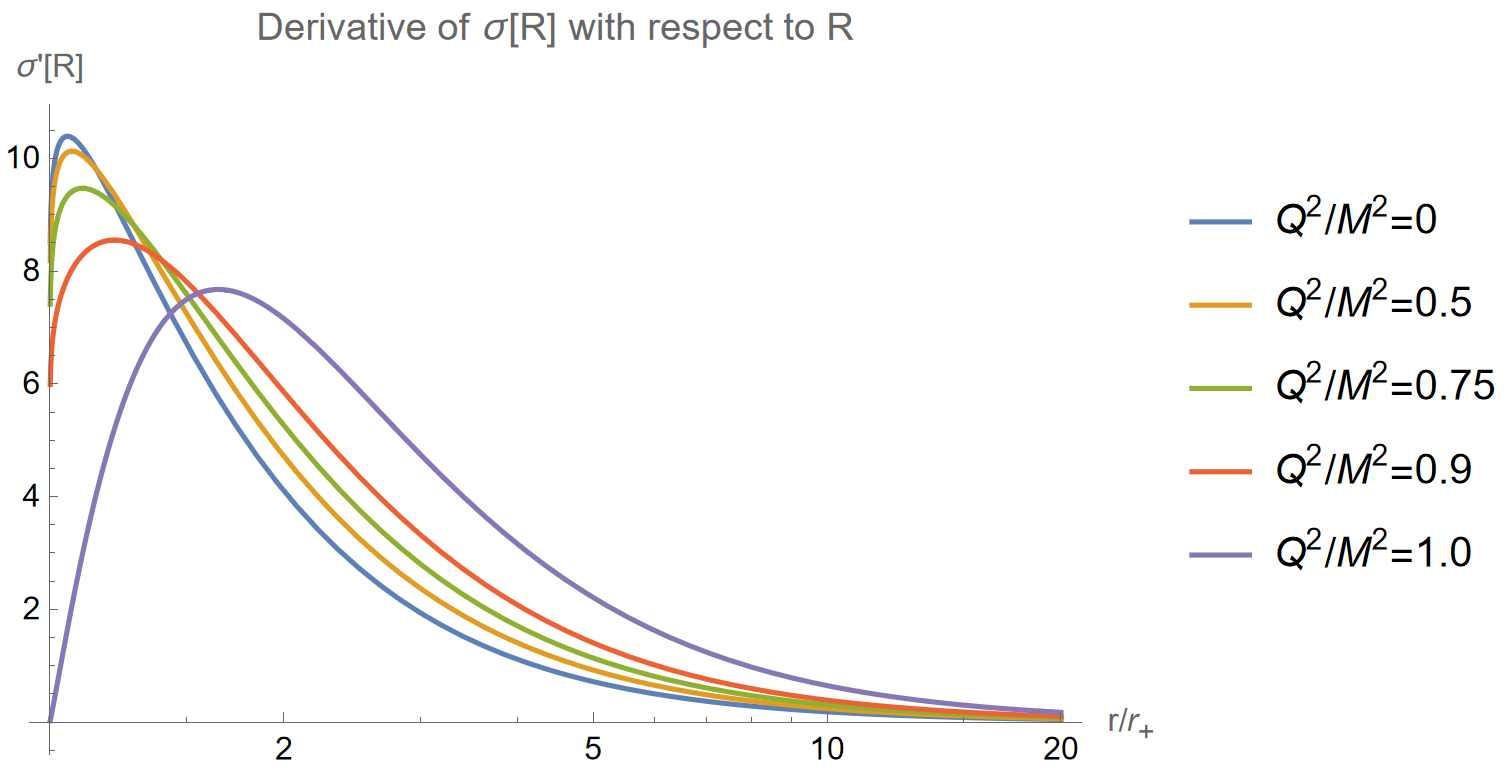}
    \caption{Graph of $\sigma^{\prime}\left(R\right)$ (the derivative of $\sigma\left(R\right)$ with respect to $R$) as a function of $r/r_{+}$, with $\lambda = 10^{-2}$ and $\ell = 10^{3}$.}
    \label{fig:sigmaderivRN}
\end{figure}
\noindent Finally, we verify the assumption that $\sigma$ can be expanded perturbatively in powers of $M^{-2}$, as in Eqn. \ref{sigmaexpansion}. This means that $\lvert \sigma_{2}/M^{2} \rvert \ll \sigma_{0}$, or equivalently, $\lvert \sigma_{2} \rvert/\sigma_{0} \ll M^{2}$. From Figures \ref{fig:sigma0graphRN} and \ref{fig:sigma2graphRN}, it is easy to see that $\sigma_{2} < \sigma_{0}$ in most of the space outside the event horizon. Thus, $\sigma_{2}/\sigma_{0} \ll M^{2}$ almost everywhere outside the event horizon. Having verified all three assumptions, we conclude that the perturbative analysis is valid almost everywhere outside the event horizon.

\end{document}